\documentclass[11pt]{article}
\pdfoutput=1
\newcommand{\VUD}{|V_{ud}|}
\newcommand{\VUS}{|V_{us}|}
\newcommand{\VUB}{|V_{ub}|}
\newcommand{\VCD}{|V_{cd}|}
\newcommand{\VCS}{|V_{cs}|}
\newcommand{\VCB}{|V_{cb}|}
\newcommand{\VTD}{|V_{td}|}
\newcommand{\VTS}{|V_{ts}|}
\newcommand{\VTB}{|V_{tb}|}
\newcommand{\VTQ}{|V_{tq}|}

\newcommand{\pt}{$p_{\rm T}$}
\newcommand{\fb}{$\rm fb^{-1}$}

\newcommand{\ttbar}{$t\bar{t}$~}

\input epsf
\usepackage{cite}
\usepackage[dvips]{graphicx}
\usepackage{color}
\usepackage{mathtools}
\usepackage{mathbbol}
\usepackage{amsmath}
\usepackage{multicol}
\usepackage{caption}
\usepackage{bmpsize}
\usepackage{multirow}
\usepackage{calc}
\newcommand{\RN}[1]{%
  \textup{\uppercase\expandafter{\romannumeral#1}}%
}

\begin{document}
\begin{center}
\vspace*{10mm}

\vspace{1cm}
{\huge \bf  Model-independent constraints on the CKM matrix elements $\VTB$, $\VTS$ and $\VTD$   } \\
\vspace{1cm}

{ \bf  Wenxing Fang$^{1}$, Barbara Clerbaux$^2$,  Andrea Giammanco$^3$ and Reza Goldouzian$^2$    }

 \vspace*{10mm}
{\emph{$^1$ School of Physics and Nuclear Energy Engineering, Beihang University, Beijing, China} }\\
   {\emph{$^2$ Interuniversity Institute for High Energies (IIHE),
Physique des particules \'el\'ementaires, Universit\'e Libre de Bruxelles, ULB, 1050, Brussels, Belgium} } \\
{\emph{$^3$ Centre for Cosmology, Particle Physics and Phenomenology (CP3), Universit\'e catholique de Louvain, Chemin du Cyclotron 2, 1348 Louvain-La-Neuve, Belgium} }
\vspace*{10mm}

\end{center}


\begin{abstract}
\noindent Single top quark production cross sections at hadron colliders are traditionally used to extract the modulus of the $V_{tb}$ element of the Cabibbo-Kobayashi-Maskawa matrix under the following assumption: $\VTB \gg \VTD, \VTS$.
For the first time, direct limits on $|V_{td}|$ and $|V_{ts}|$ are obtained using experimental data without the assumption of the unitarity of the CKM matrix. Limits on the $|V_{td}|$, $|V_{ts}|$ and $|V_{tb}|$ are extracted from differential measurements of single top quark cross sections in $t$-channel as a function of the rapidity and transverse momentum of the top quark and the light jet recoiling against the top quark.
We have shown that the pseudorapidity of the forward jet in the single top production is one of the most powerful observables for discriminating between the $|V_{td}|$ and $|V_{tb}|$ events.
We perform a global fit of top quark related CKM elements to experimental data from the LHC Runs $\RN{1}$ and $\RN{2}$ and Tevatron. Experimental data include inclusive and differential single top cross sections in $t$-channel, inclusive tW production cross section, and top quark branching ratio to b quark and W boson. We present bounds on $|V_{tb}|$, $|V_{ts}|$ and $|V_{td}|$ using current data and project the results for future LHC data sets corresponding to luminosities of 300 and 3000 \fb.
\end{abstract}

\clearpage

\section{Introduction}
\label{sec:intro}

In the Standard Model (SM) of particle physics, the elements of the Cabibbo-Kobayashi-Maskawa (CKM) matrix  \cite{Cabibbo:1963yz} are free parameters that can only be determined by experiment, with the constraints coming from the unitarity of the matrix. The values of top quark related CKM matrix elements, $\VTB$, $\VTS$ and $\VTD$, are extracted indirectly through the contribution of the top quark in B-$\bar{\text{B}}$ oscillations or loop-mediated rare K  and B meson decays \cite{Patrignani:2016xqp,Charles:2004jd}.
As the sensitivity of these measurements to the top-related elements comes from quantum loops, some model assumptions have to be imposed in their interpretation, such as the existence of only three generations of quarks and the absence of non-SM particles in the loops. The most recent global fit to the experimental results yields \cite{utfit,Charles:2015gya}:

\begin{equation}
  \left( \begin{array}{ccc}
      \VUD & \VUS  &  \VUB  \\
      \VCD & \VCS  &  \VCB  \\
     \VTD & \VTS  &  \VTB
   \end{array}  \right)    =
  \left( \begin{array}{ccc}
      0.97431 \pm 0.00015 & 0.22512 \pm 0.00067  &  0.00365 \pm 0.00012  \\
      0.22497 \pm 0.00067 & 0.97344 \pm  0.00015  &  0.04255 \pm 0.00069  \\
      0.00869 \pm 0.00014 & 0.04156 \pm 0.00056  &  0.999097 \pm 0.000024
   \end{array}  \right) \,.
  \label{eq:ckm}
\end{equation}

The value of the CKM matrix element $\VTB$ is determined under the aforementioned assumptions, but independent measurements of the top-related CKM element are needed to avoid complete reliance on the validity of the SM. In many extensions of the SM, the unitarity of the $3\times 3$ CKM matrix can be violated through the mixing of a fourth generation of quarks with the other three generations, or by non-universality of the quark couplings in electroweak interactions \cite{Alwall:2006bx}.

Vector-like quarks (VLQ)  are present in many extensions of the SM that allow sizable modifications of the CKM matrix through mixing with the SM quark families \cite{Cacciapaglia:2015ixa,Cacciapaglia:2011fx}. The new heavy VLQ that are partners of the top and bottom quarks have different gauge coupling with respect to the SM quarks. However, they can couple to the SM Higgs field through the Yukawa interactions. Consequently, the mass eigenstates  will be a mixture of the SM quarks and VLQ after spontaneous symmetry breaking of the electroweak gauge group and the unitarity of the 3$\times$3 CKM matrix will be violated. Both ATLAS and CMS Collaborations have devoted a considerable effort to search for direct production of the VLQ \cite{twikiATLAS,twikiCMS}. On the other hand, precise measurements of the   $\VTB$, $\VTS$ and $\VTD$ elements of the CKM matrix, independent of the unitarity assumption, could be used to set indirect constrains on this type of new particles~\cite{Cacciapaglia:2015ixa,Cacciapaglia:2011fx}.

The first determination of the quark mixing matrix element $\VTB$ independent of assumptions on the  $3\times 3$ CKM matrix
 unitarity was extracted from electroweak loop corrections, in particular those affecting the process $Z\rightarrow b\bar b$, in Ref.~\cite{Swain:1997mx}, yielding $\VTB = 0.77^{+0.18}_{-0.24}$  from a combination of electroweak data  from  LEP,  SLC,  Tevatron, and  neutrino  experiments.

 At hadron colliders, such as Tevatron and LHC,  top-antitop pair production via strong interactions is  the main source of top quarks. In most of the \ttbar  cross section measurements, the top quark branching ratio into a $b$ quark and a $W$ boson ($R$) is assumed to be 1. In the SM, one can write this branching ratio using the values of equation \ref{eq:ckm} as~\cite{Patrignani:2016xqp}:
 \begin{equation}
R=\frac{|V_{tb}|^{2}}{|V_{td}|^{2}+|V_{ts}|^{2}+|V_{tb}|^{2}}=0.99830^{+0.00004}_{-0.00009} \,.
\end{equation}
In Ref.~\cite{Abazov:2008yn}, the first simultaneous measurement of $R=$ BR($t \rightarrow Wb$)/$\sum_{q}$BR($t \rightarrow Wq$) and of the \ttbar  cross section, where q can be b, s or d quark, was performed in  the lepton+jets channel (one $W$ boson decays  into a quark and an anti-quark and the other
$W$ boson into a charged lepton and a neutrino) by the D0 Collaboration, obtaining $R = 0.97^{+0.09}_{-0.08}$.
This measurement was improved by using additional  data and including the dilepton channels (where both $W$ bosons decay into a charged lepton and a neutrino), yielding to $R = 0.90^{+0.04}_{-0.04}$~\cite{Abazov:2011zk}. The same measurements from the CDF experiment led to $R = 0.94^{+0.09}_{-0.09}$ and $R = 0.87^{+0.07}_{-0.07}$ for lepton+jet and dilepton channels, respectively \cite{Aaltonen:2013luz,Aaltonen:2014yua}. Top-quark pair events were also used by the CMS Collaboration to measure $R$ in the dilepton channel, obtaining $R = 1.014^{+0.032}_{-0.032}$~\cite{Khachatryan:2014nda}.
In these measurements, the R value is derived from the relative number of \ttbar events observed in b-tagged jet multiplicity distribution. 

\begin{figure}[t]
  \begin{center}
    \begin{tabular}{ccc}
      \includegraphics[width=0.3\textwidth]{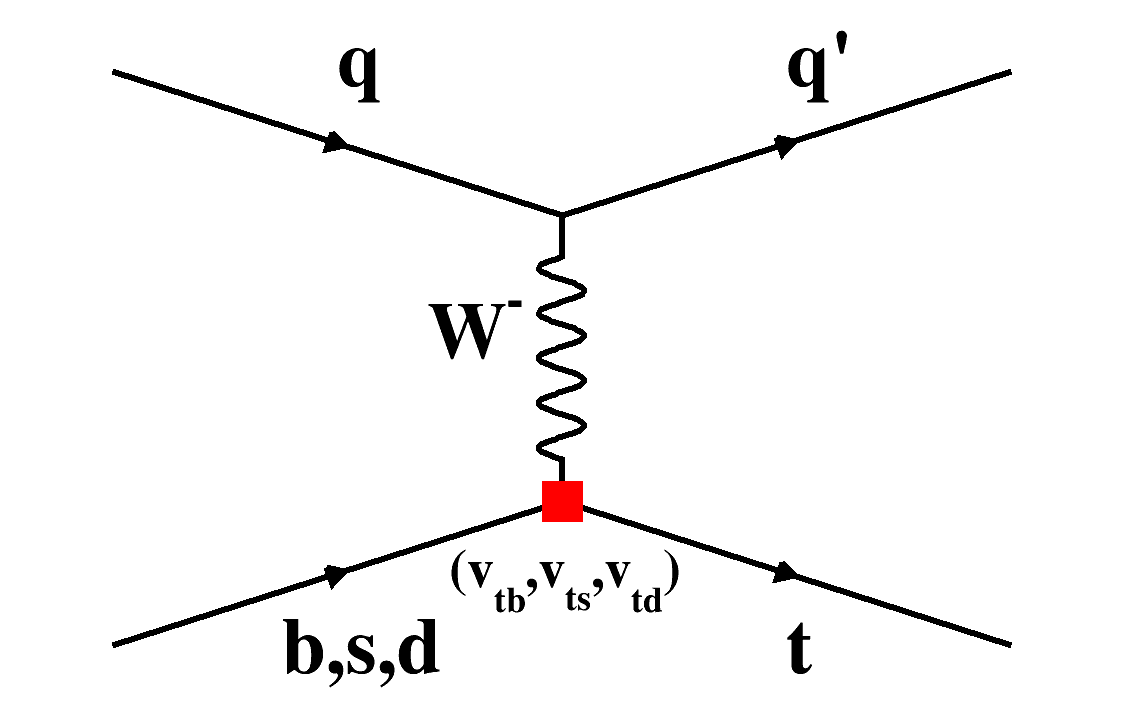}&
      \includegraphics[width=0.3\textwidth]{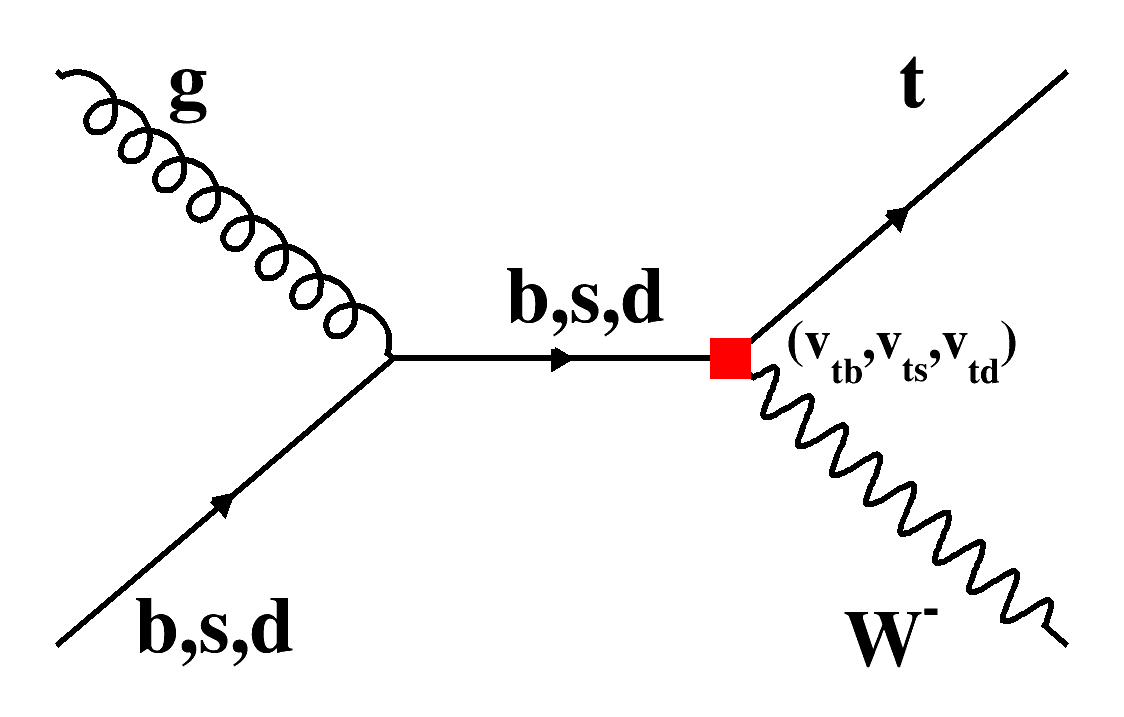}&
      \includegraphics[width=0.3\textwidth]{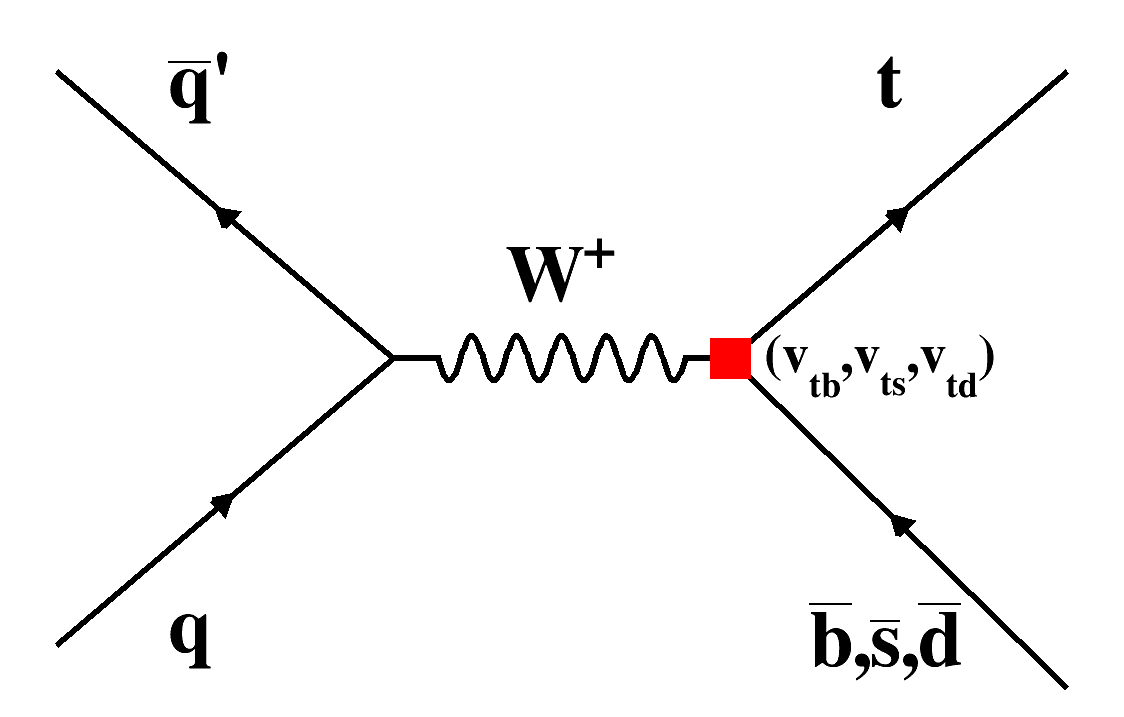}
    \end{tabular}
    \caption{ Feynman diagrams for single top quark production in $t$-channel (left), $tW$ (middle) and $s$-channel (right) modes in the presence of non-zero $\VTB$, $\VTS$ and $\VTD$ values.
    \label{Feyn}}
  \end{center}
\end{figure}

Electroweak production of top quarks provides a complementary way to probe the top quark related CKM elements, exploiting top-quark production in addition to its decay.
In the SM, single top quark production mostly proceeds via three mechanisms: the $t$-channel mode, the production of top quark in association with a $W$ boson ($tW$ mode), and the $s$-channel mode. In Figure \ref{Feyn}, the lowest level
Feynman diagrams for $t$-channel, $tW$ and $s$-channel are shown in the five-flavor scheme (i.e., considering all quarks up to the $b$ as incoming partons), without the usual approximation of ignoring $\VTS$ and $\VTD$. At the Tevatron, $t$- and $s$-channel production differ by roughly a factor of two (with the $t$-channel being the largest), while the $tW$ production cross section can be neglected for our purposes. At the LHC, the $t$-channel mode has the largest cross section followed by $tW$ while the $s$-channel production has a small cross section.
 The usage of single top quark production processes to extract direct information on $\VTD$ and $\VTS$ is discussed in Refs. \cite{Lacker:2012ek,Alwall:2006bx,Cao:2015qta}.

The  cross sections for $t$- and $s$-channel single top and anti-top quark production are expected (and observed, in the $t$-channel case) to differ by almost a factor of two at the LHC.
In the $t$-channel case this is intuitively understood by observing that, in the diagram of Figure~\ref{Feyn} (left), the incoming  quark $q$ must be $u$, $c$, $\bar{d}$, or $\bar{s}$  in order to create a top quark, whereas anti-top  production requires $q = d, s, \bar{u}$ or $\bar{c}$. Since the LHC is a $pp$ collider and the valence composition of the proton is $uud$, more top quarks are produced than anti-top quarks. As the partonic collision happens at large values of Bjorken $x$, where valence quarks dominate over the ``sea'' quarks, the charge asymmetry observable defined as:
\begin{eqnarray}\label{Eq:rt}
R_{t} = \frac{\sigma_{t-channel}^{top}}{\sigma_{t-channel}^{anti-top}}
\end{eqnarray}
gets a value close to two.
Similar considerations apply to $s$-channel production, but we will not elaborate on that case as the smallness of its cross section does not make it competitive in precision.
The value of $R_t$  can be enhanced significantly when $\VTD$ is  non-zero due to the valence-valence quark interaction in the diagram of Figure~\ref{Feyn} (left). Therefore, $R_t$ can complement the inclusive cross section for constraining $\VTD$.
In the $tW$ case, the SM predicts complete charge symmetry when $\VTD=0$, as can be understood from Figure~\ref{Feyn} (middle) noting that there are no valence $s$ and $b$ quarks.

Due to the fact that single top quark production in the $t$-channel and $tW$ are initiated by a valence quark when $\VTD$ is non-zero, and valence quarks have a very different Bjorken-$x$ spectrum with respect to sea quarks, the kinematic distributions of final state particles are different compared to the cases when a sea quark is involved (with $\VTS$ or $\VTB$) in the interaction. In Ref.~\cite{AguilarSaavedra:2010wf}, the differential distribution of the top quark rapidity is introduced as a powerful discriminator between single top $t$-channel events from non-zero Wtd and Wtb interactions. Recently, some of the differential distributions of the decay products in $tW$ have been introduced for discriminating between non-zero $\VTD$ and $\VTB$ \cite{Alvarez:2017ybk}.

In this paper we show that single top quark differential cross section measurements can be used to set more stringent limits on $\VTD$, $\VTS$ and $\VTB$ compared to the inclusive cross section measurement. We will review the most important observables for constraining the top related CKM elements directly in the differential cross section measurement of single top $t$-channel, without assuming the unitarity of the CKM matrix. We then combine experimentally measured values of the inclusive and differential single top cross sections at the LHC and Tevatron in $t$-channel and $tW$ productions with the measurements of $R$ to set constraints on $\VTD$, $\VTS$ and $\VTB$.

\section{Input observables}

For this study, several published results from the ATLAS, CMS, CDF and D0 experiments are used \cite{Aad:2014fwa,Aaboud:2017pdi,Aaboud:2016ymp,Chatrchyan:2012ep,Khachatryan:2014iya,Sirunyan:2016cdg,Aaltonen:2014mza,Abazov:2011rz,Aad:2012xca,Aad:2015eto,Aaboud:2016lpj,Chatrchyan:2012zca,Chatrchyan:2014tua,Khachatryan:2014nda,Aaboud:2017uqq,Aaltonen:2013kna,Aaltonen:2010ea}, summarized in Table~\ref{exp-results}.
All cross section measurements, inclusive and differential, are performed in a region with one $b$-tagged jet. This fact is taken into account in this paper by comparing the measured cross sections to the corresponding theoretical cross sections times $R$.

\begin{table}[ht]
\centering
\caption{The experimental observables used in this study.}
\label{exp-results}
\begin{tabular}{llll}
\hline
Dataset & $\sqrt{s}$ (TeV) & Channel (observables used in this study) & arXiv ref. \\
\hline
\hline
ATLAS   &  7              &  $t$-channel (parton level differential cross section, $R_t$)            &    1406.7844 \cite{Aad:2014fwa}        \\
ATLAS   &  8              &  $t$-channel (particle level differential cross section, $R_t$)            &    1702.02859 \cite{Aaboud:2017pdi}     \\
ATLAS   &  13             &  $t$-channel (inclusive cross section, $R_t$)            &    1609.03920 \cite{Aaboud:2016ymp}     \\
CMS   &  7                &  $t$-channel (inclusive cross section)            &    1209.4533 \cite{Chatrchyan:2012ep}     \\
CMS   &  8                &  $t$-channel (inclusive cross section, $R_t$)            &    1403.7366 \cite{Khachatryan:2014iya}     \\
CMS   &  13               &  $t$-channel (inclusive cross section, $R_t$)            &    1610.00678 \cite{Sirunyan:2016cdg}     \\
CDF   &  1.96             &  $t$-channel (inclusive cross section)            &  1410.4909 \cite{Aaltonen:2014mza}           \\
D0    &  1.96             &  $t$-channel (inclusive cross section)            &  1105.2788 \cite{Abazov:2011rz}           \\
ATLAS   &  7              &  $tW$ (inclusive cross section)            &    1205.5764 \cite{Aad:2012xca}        \\
ATLAS   &  8              &  $tW$ (inclusive cross section)            &    1510.03752 \cite{Aad:2015eto}     \\
ATLAS   &  13             &  $tW$ (inclusive cross section)            &    1612.07231 \cite{Aaboud:2016lpj}     \\
CMS   &  7                &  $tW$ (inclusive cross section)            &    1209.3489 \cite{Chatrchyan:2012zca}     \\
CMS   &  8                &  $tW$ (inclusive cross section)            &    1401.2942 \cite{Chatrchyan:2014tua}     \\
CMS   &  8                &  t$\bar{\text{t}}$ ($R$)        &    1404.2292 \cite{Khachatryan:2014nda}     \\ \hline
\end{tabular}
\end{table}

Table~\ref{xs} provides the inclusive cross sections for top quark and anti-quark production in $t$-channel and $tW$ single top quark processes as a function of $\VTD$, $\VTS$ and $\VTB$, for various center of mass energies \cite{Berger:2017zof, Gao:2017goi, Berger:2016oht, Brucherseifer:2014ama}.
We calculated these cross sections using the {\sc MadGraph\_}a{\sc mc@NLO}v2.6.0 framework \cite{Alwall:2014hca} in the five flavor scheme, at NLO for $t$-channel and LO for $tW$ production; in the latter case, a constant $k$-factor of 1.26 is then applied in the subsequent analysis, assumed to be independent of $\VTD$, $\VTS$ and $\VTB$~\cite{Kidonakis:2015nna}.
The proton parton distribution functions (PDF) and their uncertainties are evaluated with the reference sets of NNPDF3.1 \cite{Ball:2014uwa}. The mass of the top quark is set to
$m_t$ = 172.5 GeV and the factorization and renormalisation scales are fixed to the top quark mass.

\begin{table}[ht]
\caption{Single top quark production cross section (in pb) via $t$-channel and $tW$-channel  at center of mass energies of 7, 8 and 13 TeV of the $pp$ collision data and 1.96 TeV of the $p\bar{p}$ collision data as a function of $\VTD$, $\VTS$ and $\VTB$. No phase space cut is implemented on the final-state particles.  NNPDF3.1 is used to evaluate the parton densities,
while the renormalization scale and factorization scale are set to the top quark mass ($m_t$ = 172.5 GeV).
}
\label{xs}
\centering
\resizebox{\linewidth}{!}{%
\begin{tabular}{lllll}
\hline
Collision - $\sqrt{s}$ (TeV) & Channel  & Cross section (pb) & precision  \\
\hline
\hline
$pp$ - 7              &  $t$-channel (top)            &   $43.7\VTB^2 + 104.2\VTS^2 + 329.2\VTD^2 $  & NLO        \\
$pp$ - 7              &  $t$-channel (anti-top)       &  $22.8\VTB^2 + 52.2\VTS^2 + 84.4\VTD^2 $   & NLO      \\ \hline
$pp$ - 8              &  $t$-channel (top)            &  $56.3\VTB^2 + 132.5\VTS^2 + 406.4\VTD^2 $  &NLO       \\
$pp$ - 8              &  $t$-channel (anti-top)       &  $30.7\VTB^2 + 69.6\VTS^2 + 109.0\VTD^2 $    &NLO      \\ \hline
$pp$ - 13             &  $t$-channel (top)            &  $136.5\VTB^2 + 300.72\VTS^2 + 772.8\VTD^2 $  &NLO        \\
$pp$ - 13             &  $t$-channel (anti-top)            &  $82.1\VTB^2 + 177.1\VTS^2 + 260.4 \VTD^2 $  &NLO        \\ \hline
$p\bar{p}$ - 1.96 &  $t$-channel (top+anti-top) &  $2.1\VTB^2 + 6.3\VTS^2 + 24.3 \VTD^2 $  &NLO     \\  \hline
$pp$ - 7              &  $tW$ (top)         &  $6.3\VTB^2 + 12.2\VTS^2 +84.0 \VTD^2 $    &LO ($k$-factor = 1.26)      \\
$pp$ - 7              &  $tW$ (anti-top)         &  $6.3\VTB^2 + 12.2\VTS^2  + 22.6\VTD^2 $ &LO ($k$-factor = 1.26)          \\\hline
$pp$ - 8              &  $tW$ (top)         &  $8.9\VTB^2 + 16.9\VTS^2 +109.1 \VTD^2 $   &LO  ($k$-factor = 1.26)       \\
$pp$ - 8              &  $tW$ (anti-top)         &  $8.9\VTB^2 + 16.9\VTS^2 + 31.0\VTD^2 $  &LO  ($k$-factor = 1.26)        \\ \hline
$pp$ - 13             &  $tW$ (top)         &  $28.0\VTB^2 +50.6 \VTS^2 + 265.8\VTD^2 $    &LO  ($k$-factor = 1.26)      \\
$pp$ - 13             &  $tW$ (anti-top)         &  $28.0\VTB^2 +50.6 \VTS^2 + 89.0\VTD^2 $   &LO  ($k$-factor = 1.26)    \\ \hline
\end{tabular}}
\end{table}

\subsection{Fiducial and differential cross sections of single top quark production}
\label{fid}

In general, measurements of the single top  quark production cross section provide a unique testing ground for measuring the top quark related CKM elements.
Single top quark production via $t$-channel was established in $p\bar{p}$  collisions at the Tevatron \cite{Aaltonen:2014mza,Abazov:2011rz} and its cross section is measured precisely in $pp$ collisions at the LHC \cite{Aad:2014fwa,Aaboud:2017pdi,Aaboud:2016ymp,Chatrchyan:2012ep,Khachatryan:2014iya,Sirunyan:2016cdg}.
In addition to the inclusive cross section measurement of single top quark in $t$-channel,  differential cross sections are measured as a function of the top and anti-top quark transverse momentum (\pt) and the absolute value of its rapidity ($|\text{y}|$) at parton level (where the top quark 4-momentum is defined before the decay and after emission of QCD radiation) at 7 and 8 TeV by the ATLAS Collaboration \cite{Aad:2014fwa,Aaboud:2017pdi}. At 8 TeV, differential cross sections are measured also as a function of the kinematic observables of the accompanying light jet ($j$) in the $t$-channel exchange.
As it was discussed in Section~\ref{sec:intro}, the kinematic distributions of final state particles via single top quark production are distorted with respect to the SM predictions, in case of non-zero values of $\VTS$ and $\VTD$.

\begin{figure}[!ht]
  \begin{center}
    \begin{tabular}{cc}
      \includegraphics[width=0.47\textwidth]{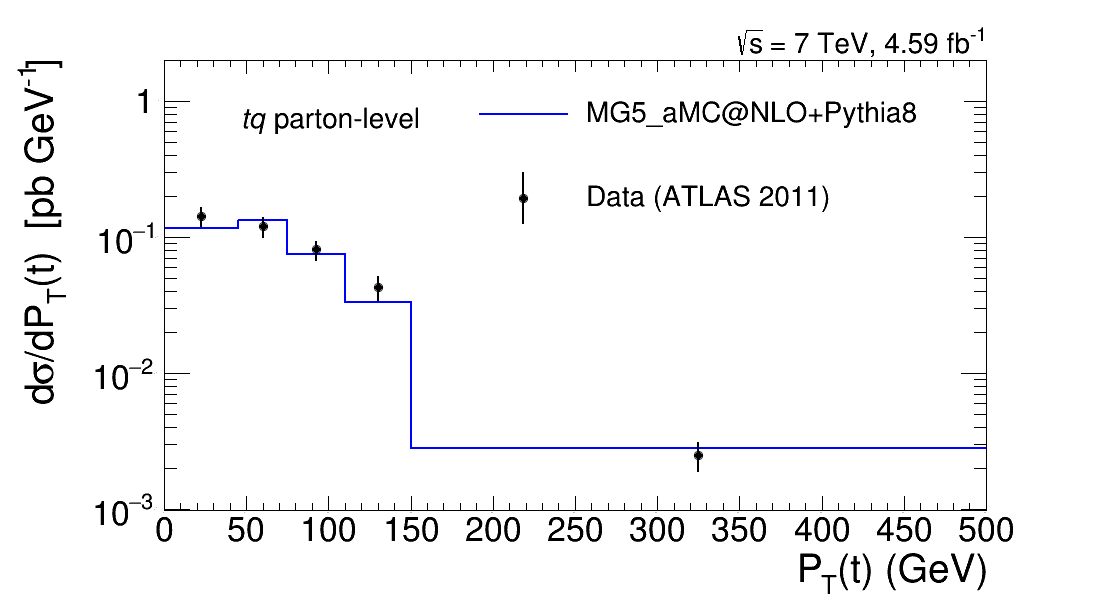}&
      \includegraphics[width=0.47\textwidth]{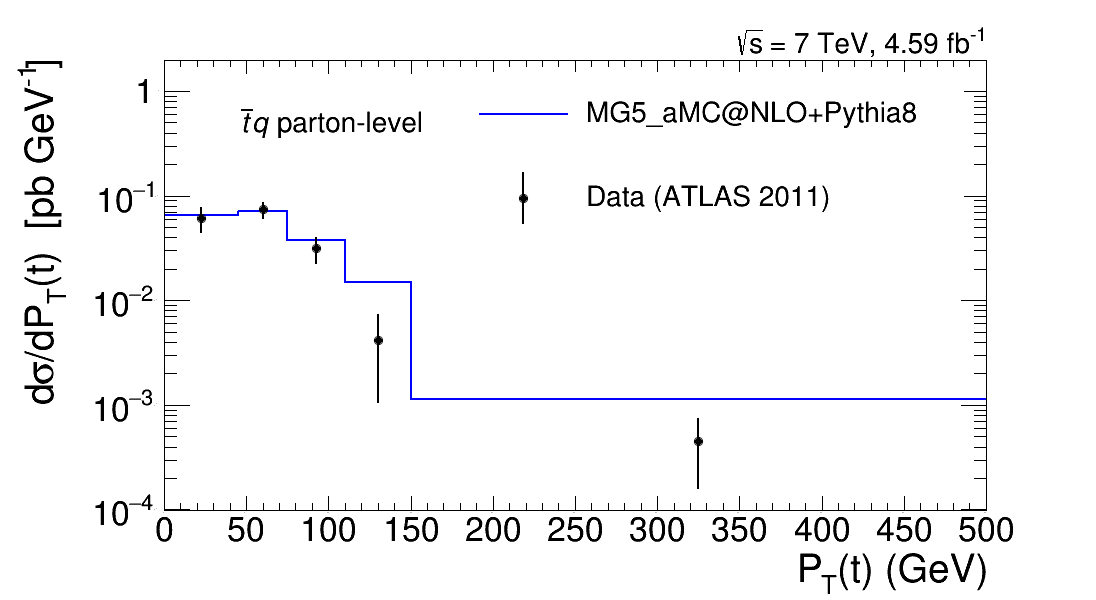}\\
      \includegraphics[width=0.47\textwidth]{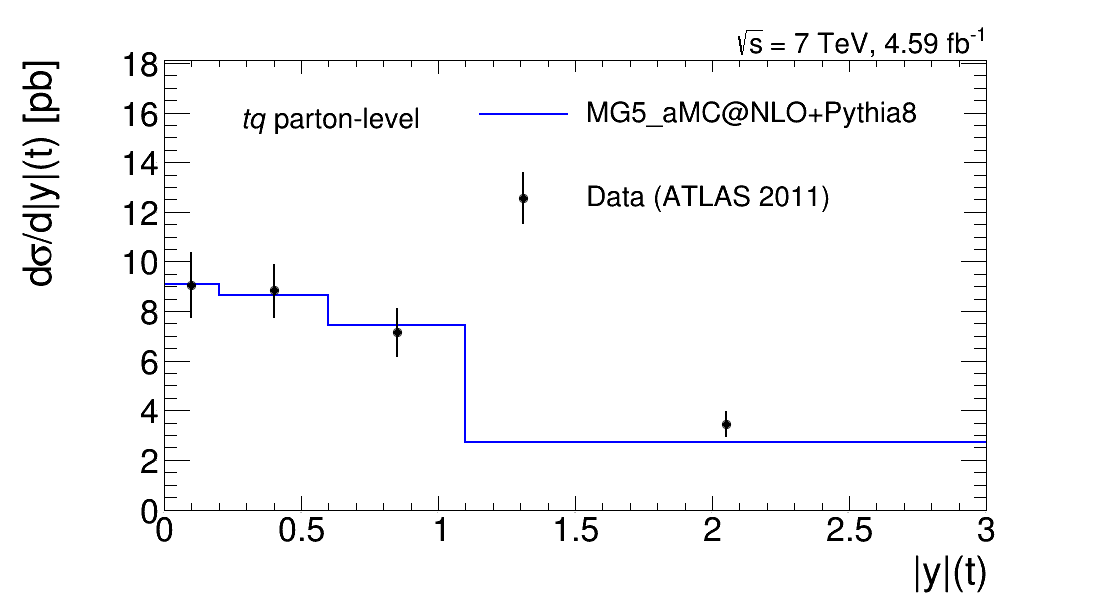}&
      \includegraphics[width=0.47\textwidth]{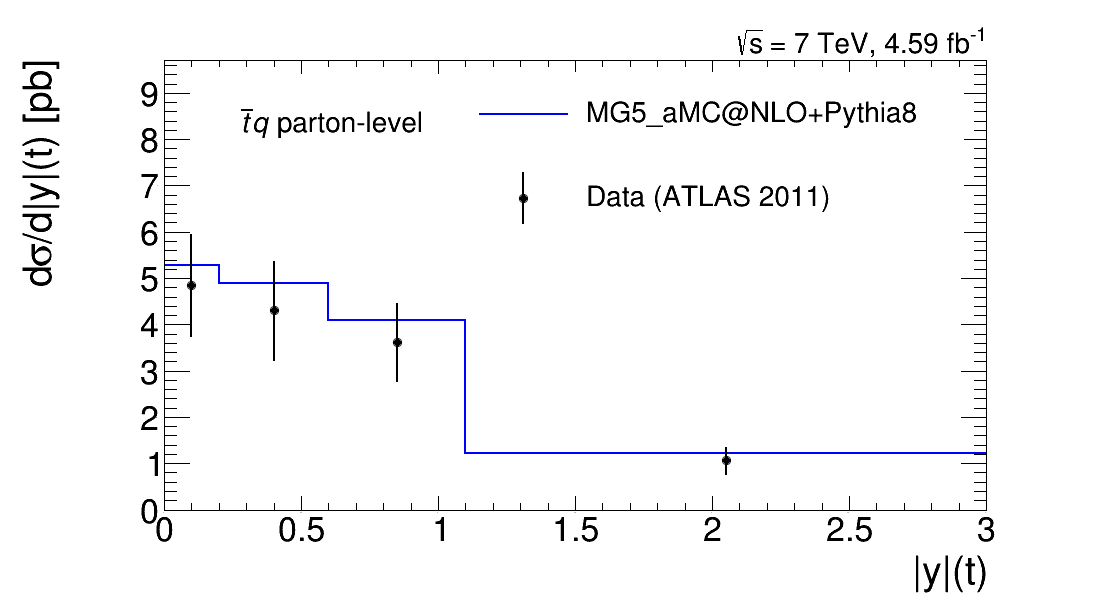}\\
    \end{tabular}
    \caption{Absolute differential single top (left) and anti-top (right) quark cross sections at parton level as a function of the top \pt~(upper row) and top $|\text{y}|$ (lower row). Data are taken from Ref. \cite{Aad:2014fwa} (ATLAS Collaboration, 7 TeV)  and are compared to the SM prediction using the MC samples generated by  {\sc MadGraph\_}a{\sc mc@NLO}  interfaced to {\sc Pythia} for showering and hadronization.
    \label{PL-Diff-7TeV}}
  \end{center}
\end{figure}

In Figure \ref{PL-Diff-7TeV}, the absolute differential $t$-channel single top (anti-top) quark cross sections  at parton  level measured by the ATLAS Collaboration at 7 TeV~\cite{Aad:2014fwa} is compared to the SM prediction using the MC samples generated by  {\sc MadGraph\_}a{\sc mc@NLO}  interfaced to {\sc Pythia} (v8.205) for showering and hadronization \cite{Sjostrand:2014zea}. These measurements are extracted under the assumptions $\VTB \approx 1$ and
$\VTD \approx \VTS \approx 0$. This implicit model dependence induces a bias when interpreting the results to extract  $\VTS$ and $\VTD$.
Therefore, we assign 5\% uncertainty on the acceptance of the parton level signal efficiency for $\VTS$ and $\VTD$ in the global fit (see Section \ref{fit}).
In addition to the parton level, differential cross sections are also reported, at particle level, within a fiducial phase space by the ATLAS Collaboration at 8 TeV \cite{Aaboud:2017pdi}. The fiducial measurements are more model independent and have smaller theory uncertainties compared to the parton level measurements.

The definition of fiducial phase space is based on simple requirements  on the particle level objects. Particle level objects are reconstructed from stable particles with a lifetime larger than $0.3 \times 10^{-10}$ s appearing  after showering and hadronization steps. We follow closely the particle level object definition  presented in \cite{Aaboud:2017pdi} by the ATLAS Collaboration for constructing the leptons, jets and $b$-jets. Events are selected if exactly one particle level muon or electron with \pt $>$ 25 GeV and $|\eta| <$ 2.5 is found. Events must contain exactly two particle level jets with \pt $>$ 30 GeV and $|\eta| <$ 4.5 while one of them be tagged as particle level $b$-jet in $|\eta| <$ 2.5 region. Based on the selected particle level lepton, $b$-jet and missing transverse energy, top quark is reconstructed as described in Ref.~\cite{Aaboud:2017pdi}.

It is worth discussing the effect of the top quark related CKM elements on the backgrounds.
The most important background in the differential $t$-channel single top quark cross section measurements that is affected by $\VTB$, $\VTS$ and $\VTD$ variations is the \ttbar process.
The contribution of the \ttbar background with two $b$-quark jets from the top and anti-top decays varies as a function of R$^2$. In addition, more \ttbar events contribute to the signal region when the top quark decays to a W boson and a $b$-quark and the anti-top quark decays to a W boson and an anti-d/s quark or vice versa. This cross section is also function of R through $\sigma$($t\bar{t}$) $\times$ R(1-R).
The effect of $\VTB$, $\VTS$ and $\VTD$ variations on the background estimation   is expected to be negligible because of the following reasons. First, R is measured precisely by the CMS experiment in a region enriched by \ttbar events, $R = 1.014^{+0.032}_{-0.032}$~\cite{Khachatryan:2014nda}. In this measurement, the R dependence  of  single top quark contribution is taken into account. Second, in Ref.~\cite{Aaboud:2017pdi}, a neural network is employed to separate single top events from \ttbar events using kinematic distributions. This  reduced the \ttbar event contribution significantly and after cutting on the MVA distribution, the \ttbar contribution is only 10\% of the selected events, while the single top is 70\% (see Table 8 of Ref.~\cite{Aaboud:2017pdi}). The \ttbar contamination is  more important in the differential cross section measurements of the $tW$-channel~\cite{Aaboud:2017qyi}, which are not considered in this study because of their insufficient precision.

\begin{figure}[!ht]
  \begin{center}
    \begin{tabular}{cc}
      \includegraphics[width=0.47\textwidth]{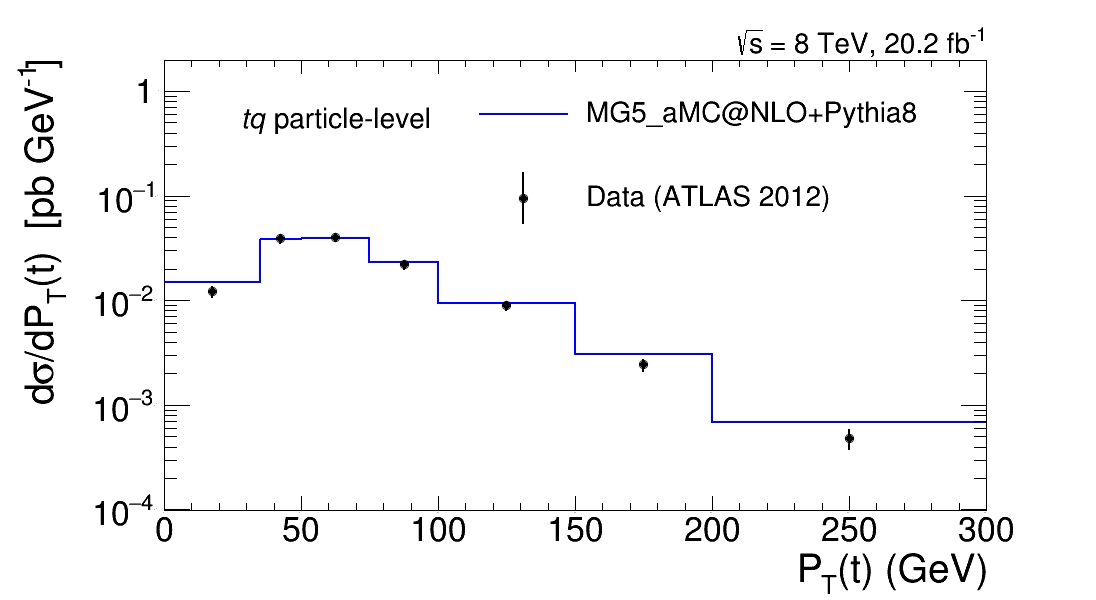}&
      \includegraphics[width=0.47\textwidth]{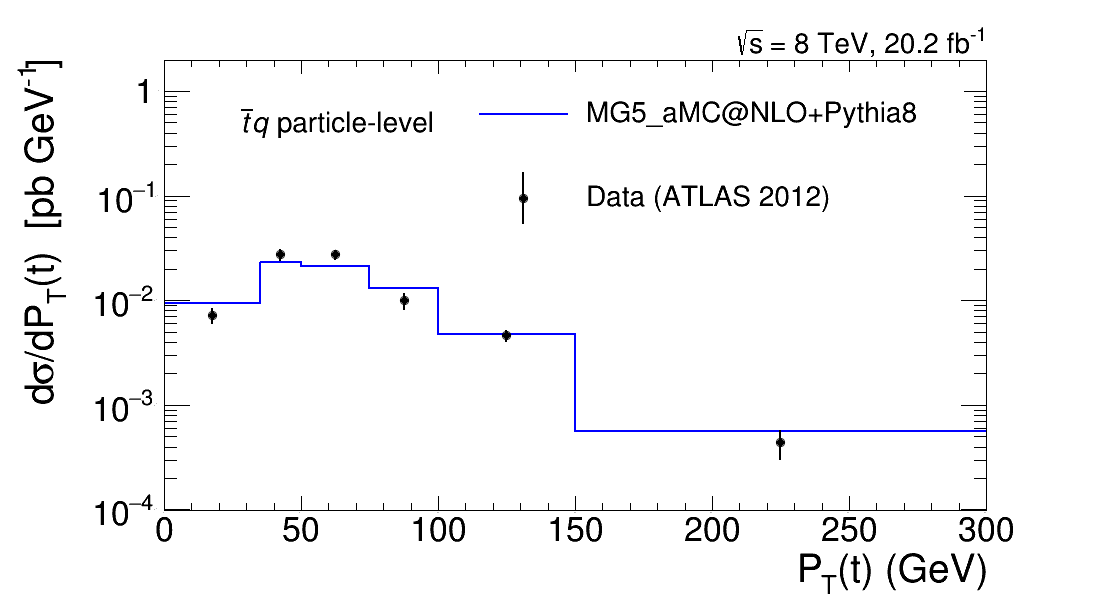}\\
      \includegraphics[width=0.47\textwidth]{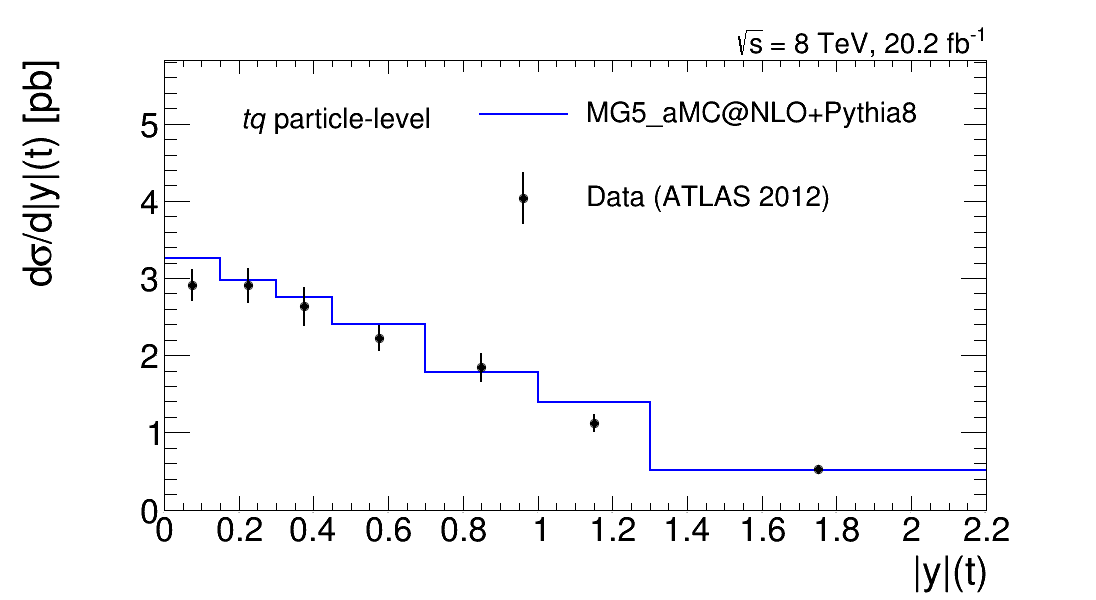}&
      \includegraphics[width=0.47\textwidth]{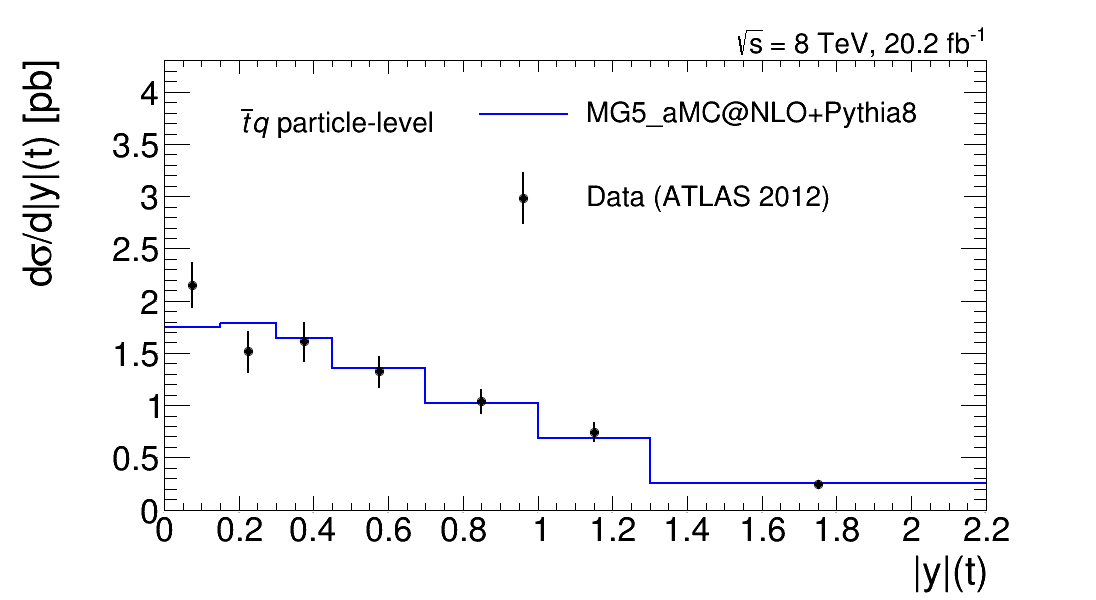}\\
      \includegraphics[width=0.47\textwidth]{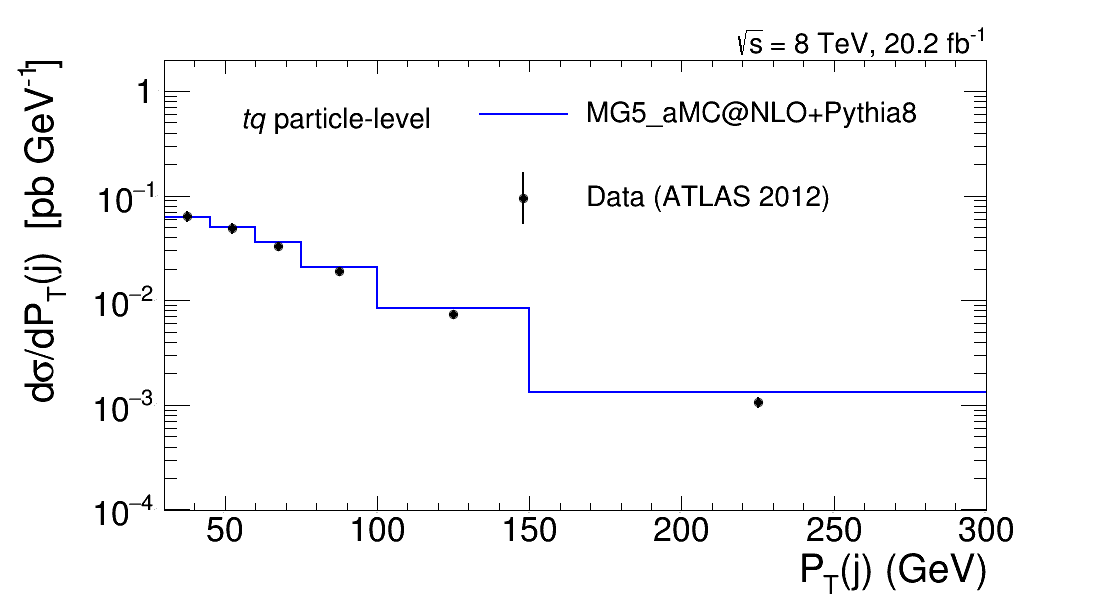}&
      \includegraphics[width=0.47\textwidth]{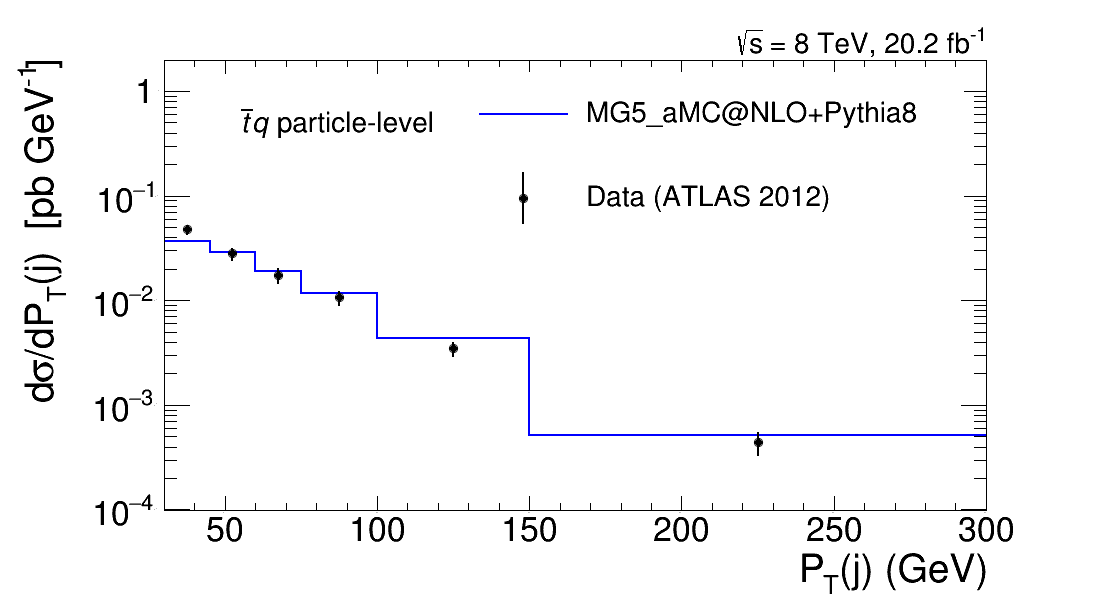}\\
      \includegraphics[width=0.47\textwidth]{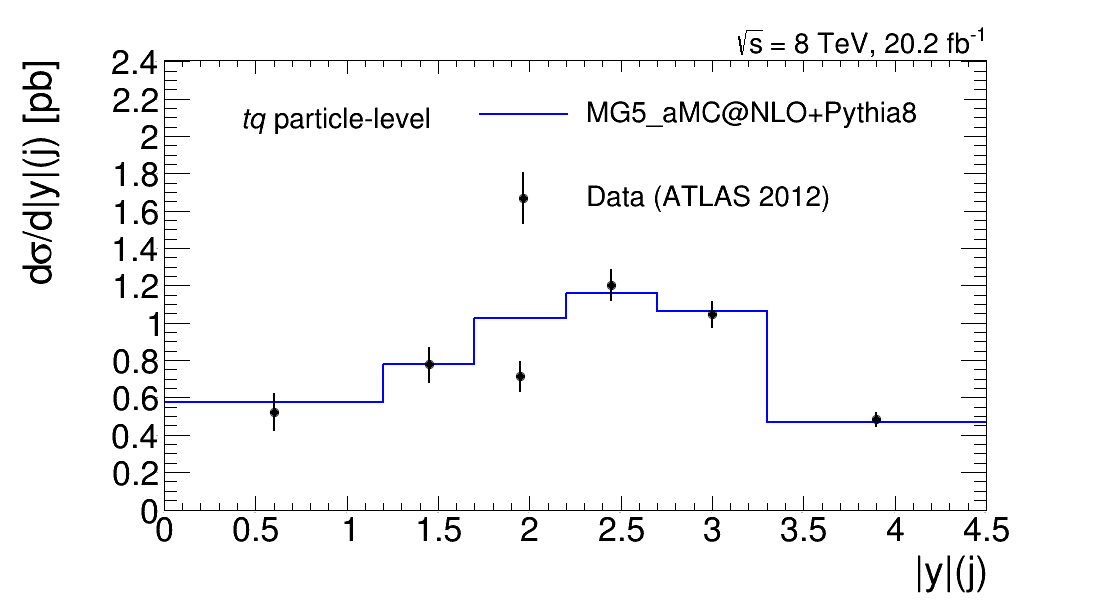}&
      \includegraphics[width=0.47\textwidth]{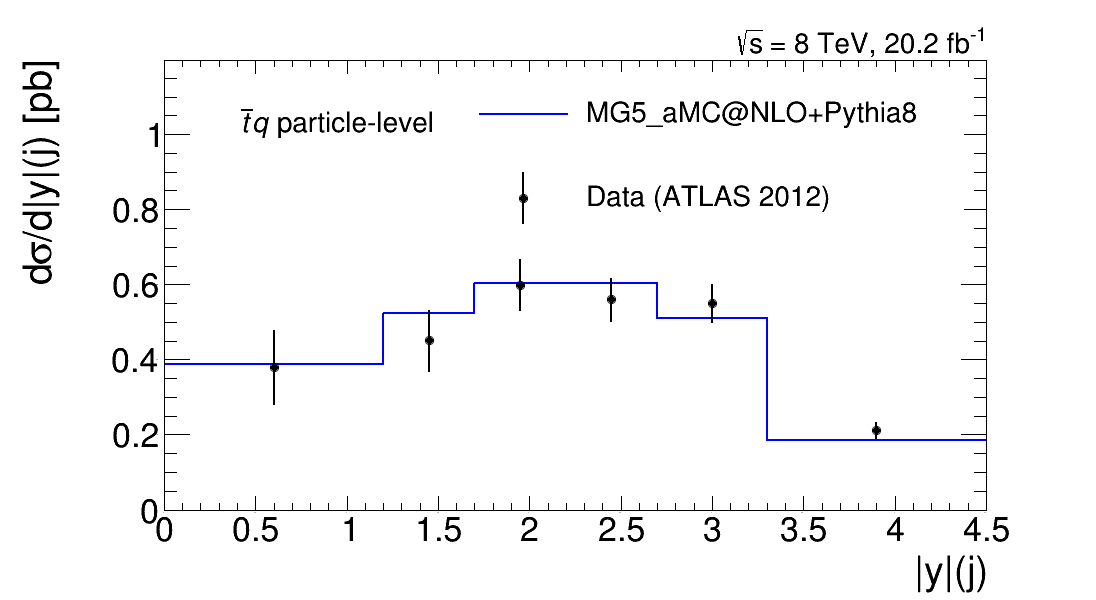}\\
    \end{tabular}
    \caption{Absolute differential single top (left) and anti-top (right) quark cross sections at particle level as a function of the top \pt~ (first row), top $|\text{y}|$ (second row),  forward jet \pt (third row) and forward jet $|\text{y}|$ (fourth row). Data are taken from Ref. \cite{Aaboud:2017pdi} (ATLAS Collaboration, 8 TeV) and are compared to the SM prediction using the MC samples generated by  {\sc MadGraph\_}a{\sc mc@NLO}  interfaced to {\sc Pythia} for showering and hadronization.
    \vspace{10 mm}
    \label{PL-Diff}}
  \end{center}
\end{figure}

Figure~\ref{PL-Diff} shows the absolute differential $t$-channel single top (anti-top) quark cross sections as a function of the top (anti-top) \pt , the rapidity $|\text{y}|$, and the forward jet \pt ~and $|\text{y}|$, measured at particle level from the ATLAS data at 8~TeV and compared to the corresponding SM predictions. With respect to Ref.~\cite{Aaboud:2017pdi}, we multiply each bin by a factor of 0.324 to take into account the branching ratio of $W$ decay into leptons.
The SM predictions at NLO in QCD come from MC samples generated with {\sc MadGraph\_}a{\sc mc@NLO} in the four-flavor scheme and normalized to the SM inclusive cross section, $\sigma_{tq}=56.3$ pb and  $\sigma_{\bar{t}q}=30.7$ pb (see Table~\ref{xs}) \cite{Campbell:2009ss}. The parton showering, hadronization and the underlying event simulation are modeled by {\sc Pythia}v8.2 using tune CUETP8M1 \cite{Khachatryan:2015pea}.
The fiducial acceptances for single top production in $t$-channel are determined to be A$_{\rm fid}^{\rm \VTD}$ (tq) = 6.9\%, A$_{\rm fid}^{\rm \VTD}$ ($\bar{\rm t}$q) = 7.5\%, A$_{\rm fid}^{\rm \VTS}$ (tq) = 7.4\%, A$_{\rm fid}^{\rm \VTS}$ ($\bar{\rm t}$q) = 7.5\%, A$_{\rm fid}^{\rm \VTB}$ (tq) = 6.0\% and A$_{\rm fid}^{\rm \VTB}$ ($\bar{\rm t}$q) = 6.1\%.
As already reported by the ATLAS Collaboration, good agreement between the measured absolute differential cross section and SM prediction ($\VTB$ = 1) is observed.

\begin{figure}[!ht]
  \begin{center}
    \begin{tabular}{cc}
      \includegraphics[width=0.47\textwidth]{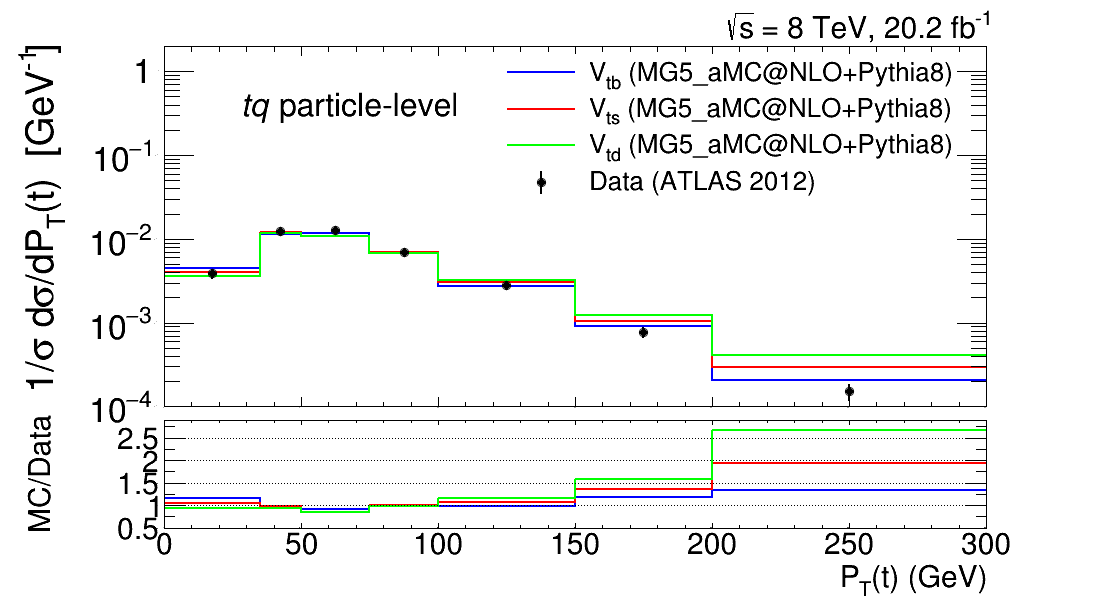}&
      \includegraphics[width=0.47\textwidth]{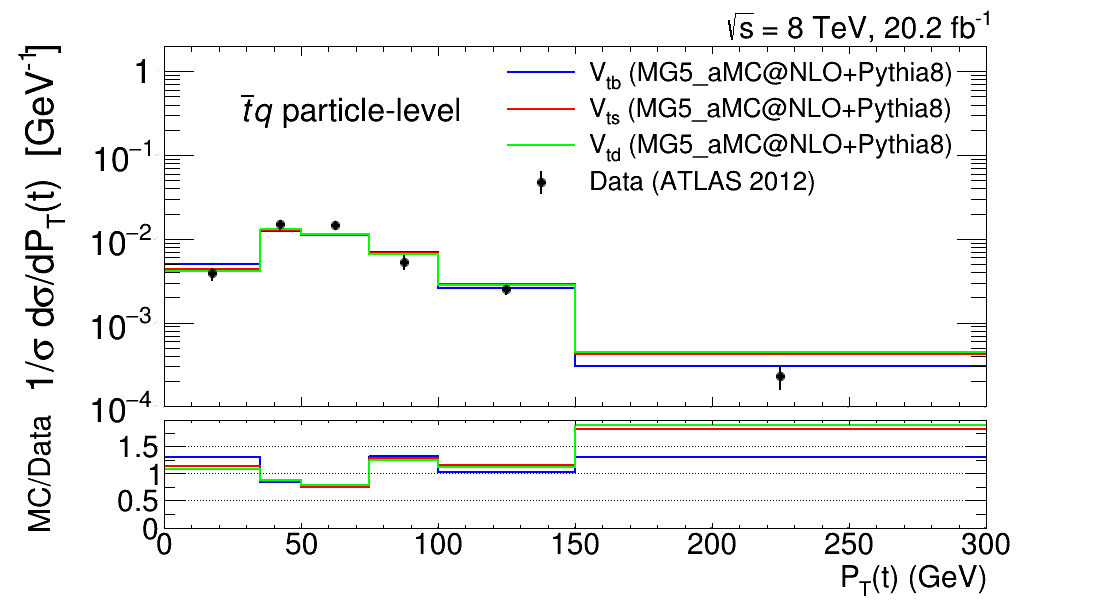}\\
      \includegraphics[width=0.47\textwidth]{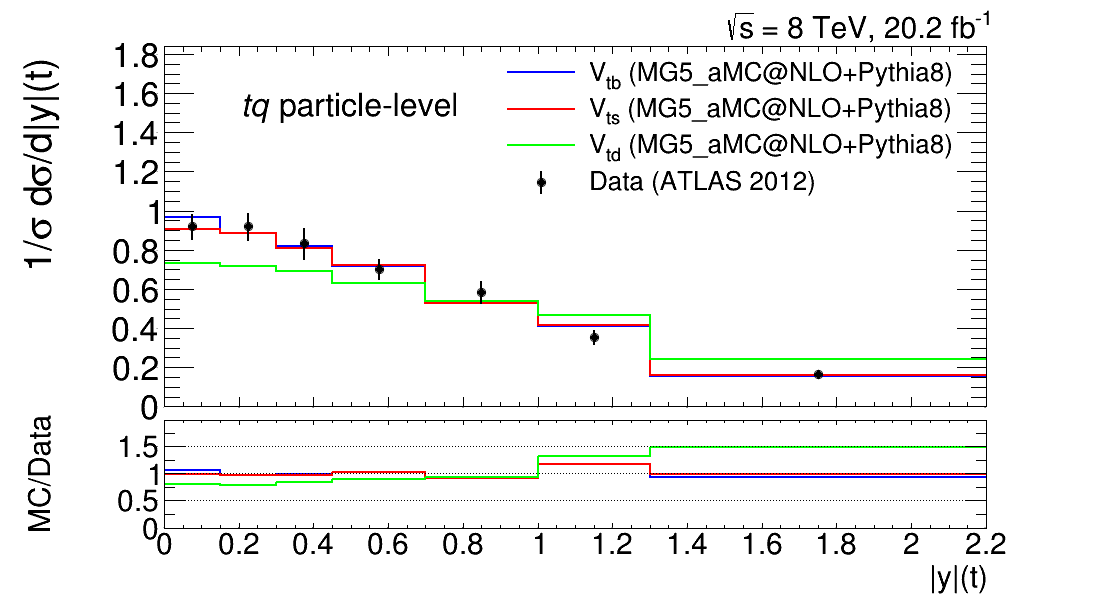}&
      \includegraphics[width=0.47\textwidth]{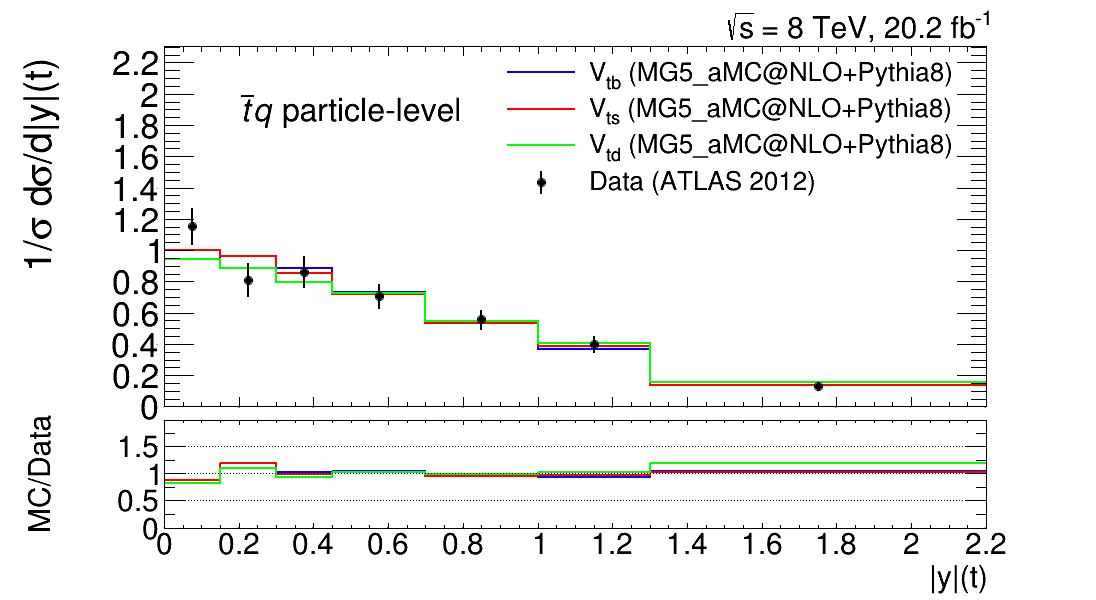}\\
      \includegraphics[width=0.47\textwidth]{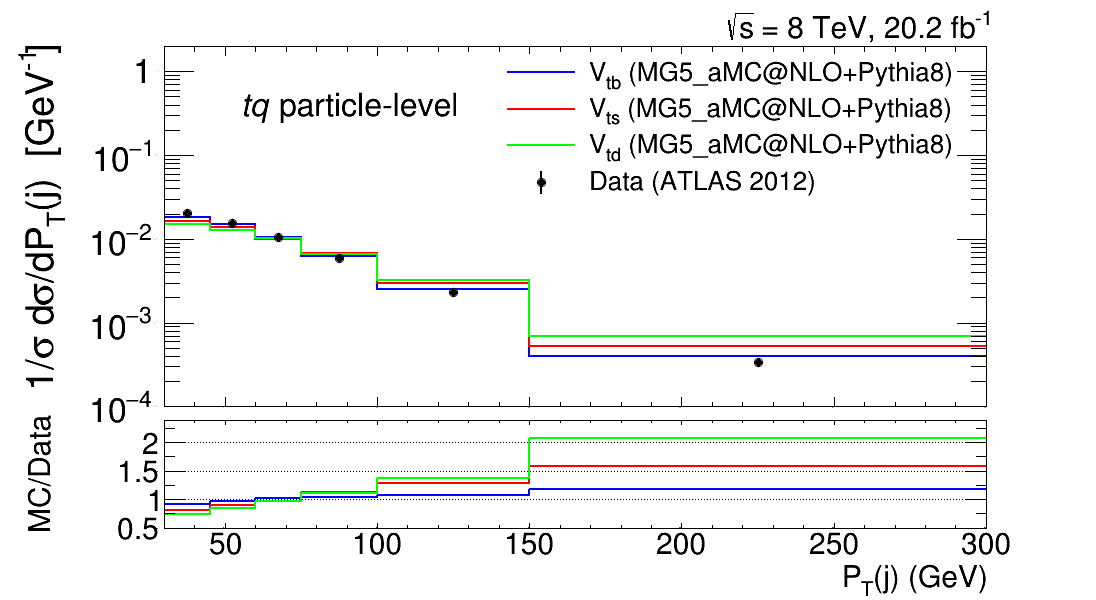}&
      \includegraphics[width=0.47\textwidth]{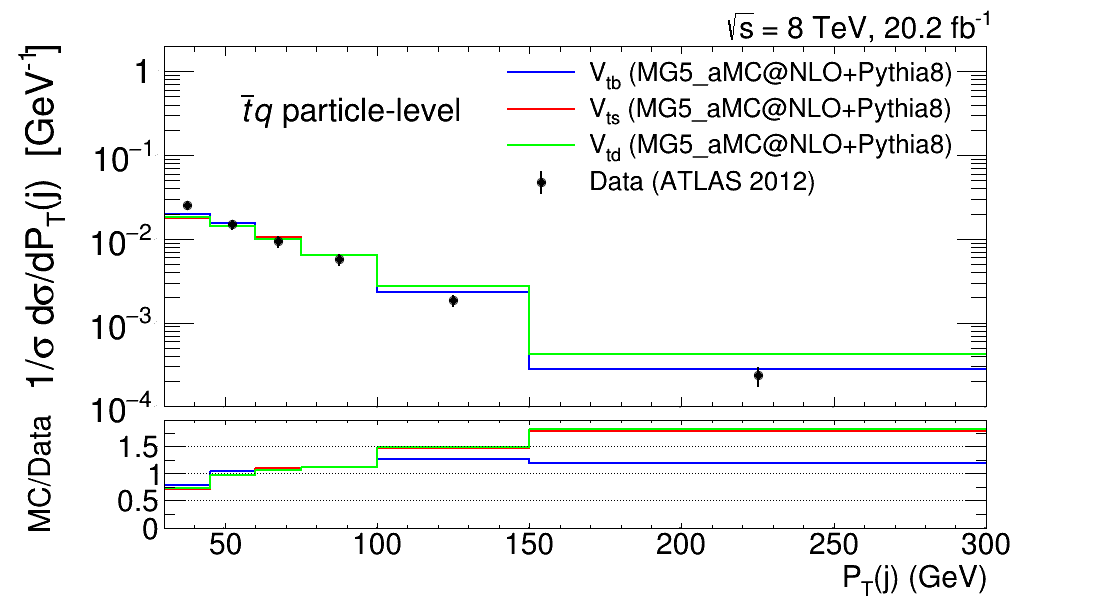}\\
      \includegraphics[width=0.47\textwidth]{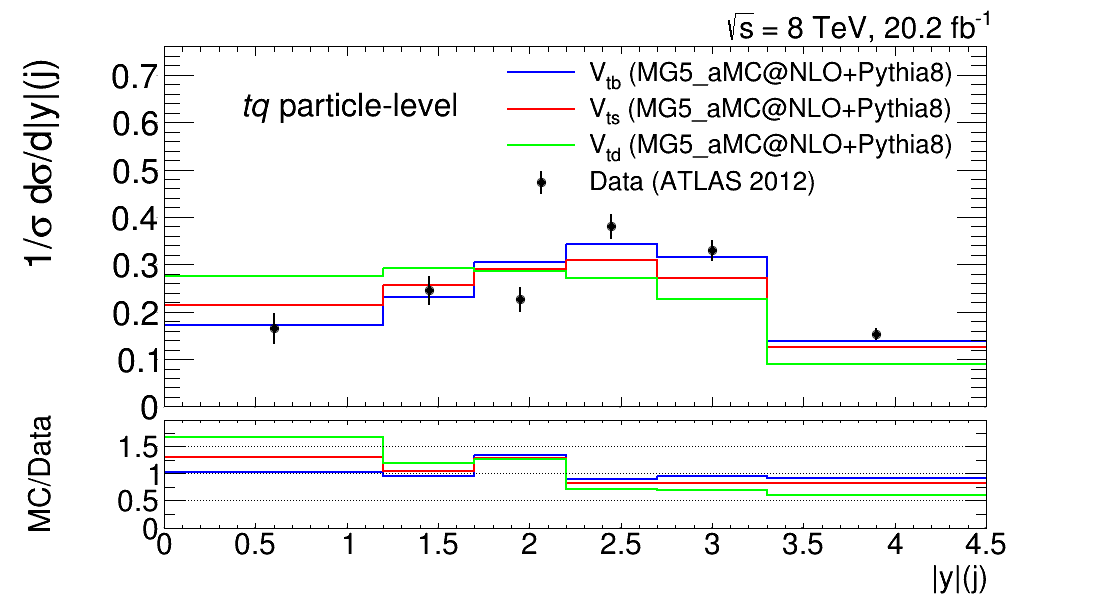}&
      \includegraphics[width=0.47\textwidth]{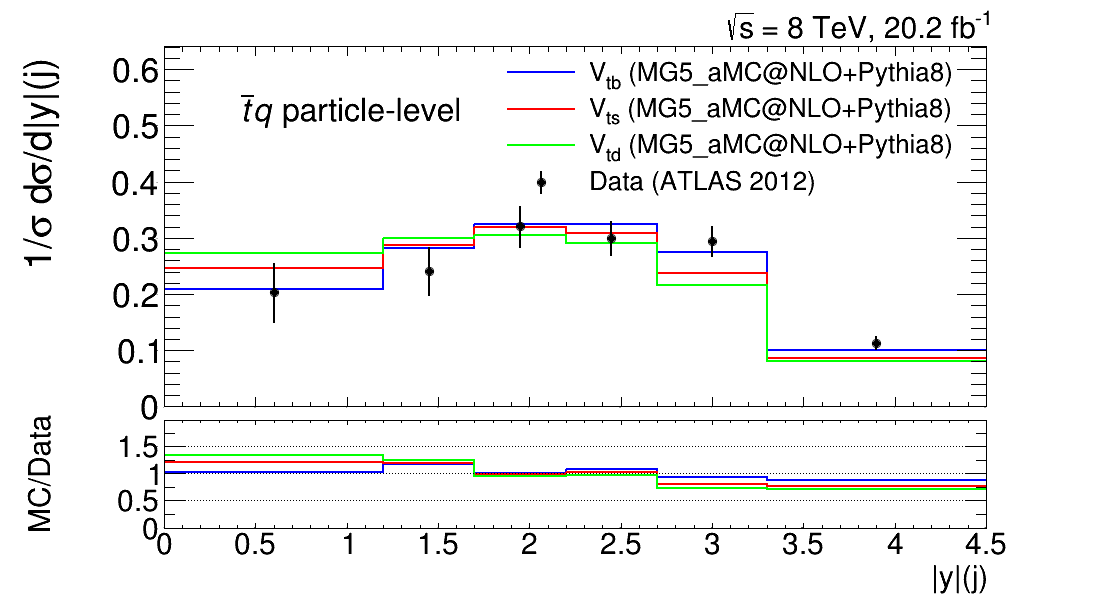}\\
    \end{tabular}
    \caption{Comparison of the normalized differential cross sections for single top (left) and anti-top (right) production via Wtb (blue line), Wts (red line) and Wtd (green line) interactions at particle level. Observables are the same as those shown in Figure \ref{PL-Diff}. Data are taken from Ref. \cite{Aaboud:2017pdi} (ATLAS Collaboration, 8 TeV).
    \label{PL-NormDiff}}
  \end{center}
\end{figure}

Normalized differential cross sections for the $t$-channel single top and anti-top quark production as a function of the \pt ~and $|\text{y}|$ of the top quark (or anti-top quark) and as a function of the  \pt ~and $|\text{y}|$ of the forward jet, at particle level, are shown in Figure \ref{PL-NormDiff}. Data are taken from Ref. \cite{Aaboud:2017pdi} and are compared to the MC predictions for single top and anti-top quark production via Wtb, Wts and Wtd interactions.
The most distinctive variable is the rapidity of the forward jet. Single top events from Wtd and Wts interactions are more central compared to the events from Wtb interactions.
Another important variable is the rapidity of the top quark followed by the \pt ~of the forward jet.
For all shown observables, the single top quark production has more sensitivity to distinguish the various production  mechanisms compared to the single anti-top production.
Although small fluctuations are observed in few bins, data follow the $\VTB = 1$ shape closely.

\begin{figure}[!ht]
  \begin{center}
    \begin{tabular}{cc}
      \includegraphics[width=0.47\textwidth]{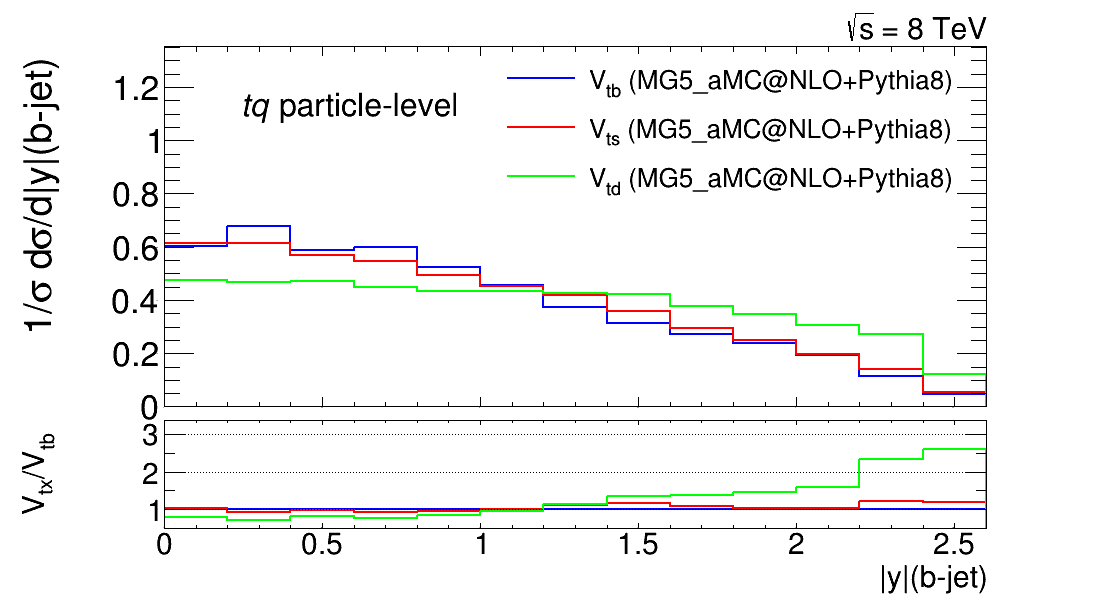}&
      \includegraphics[width=0.47\textwidth]{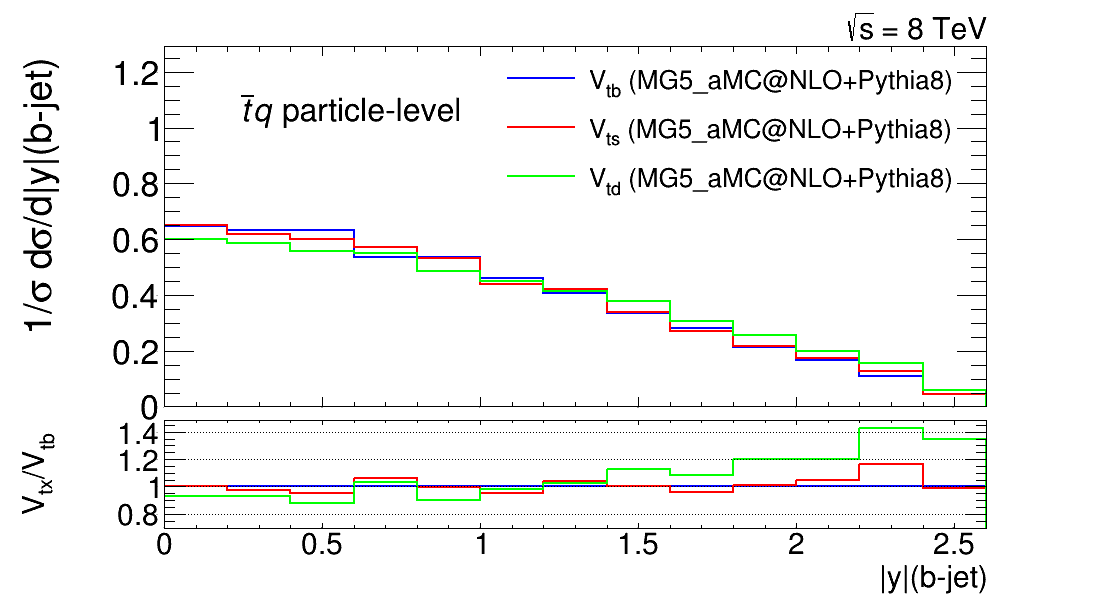}\\
      \includegraphics[width=0.47\textwidth]{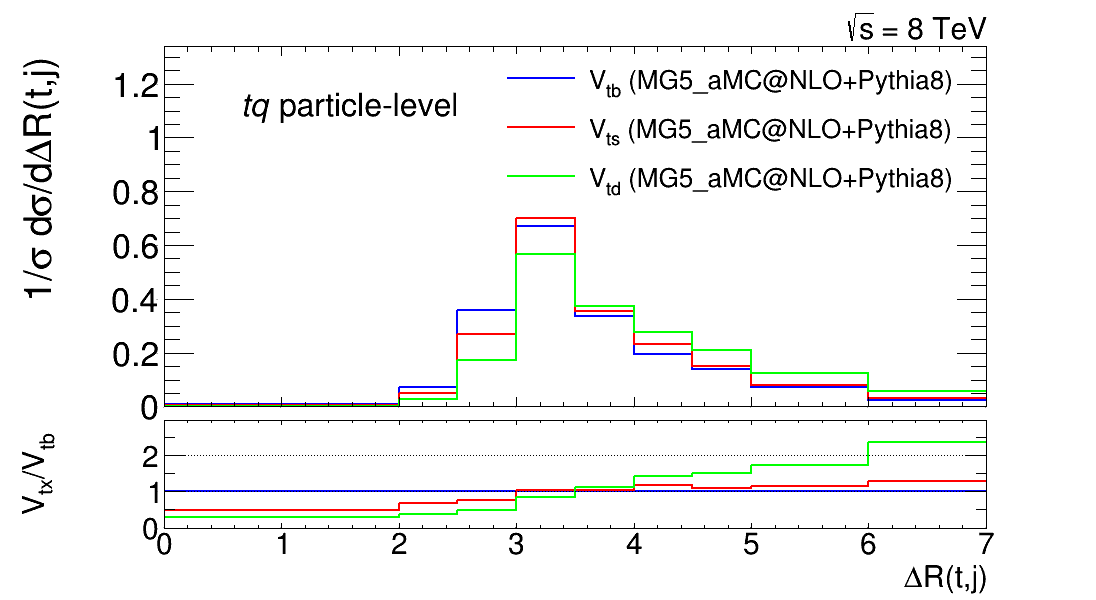}&
      \includegraphics[width=0.47\textwidth]{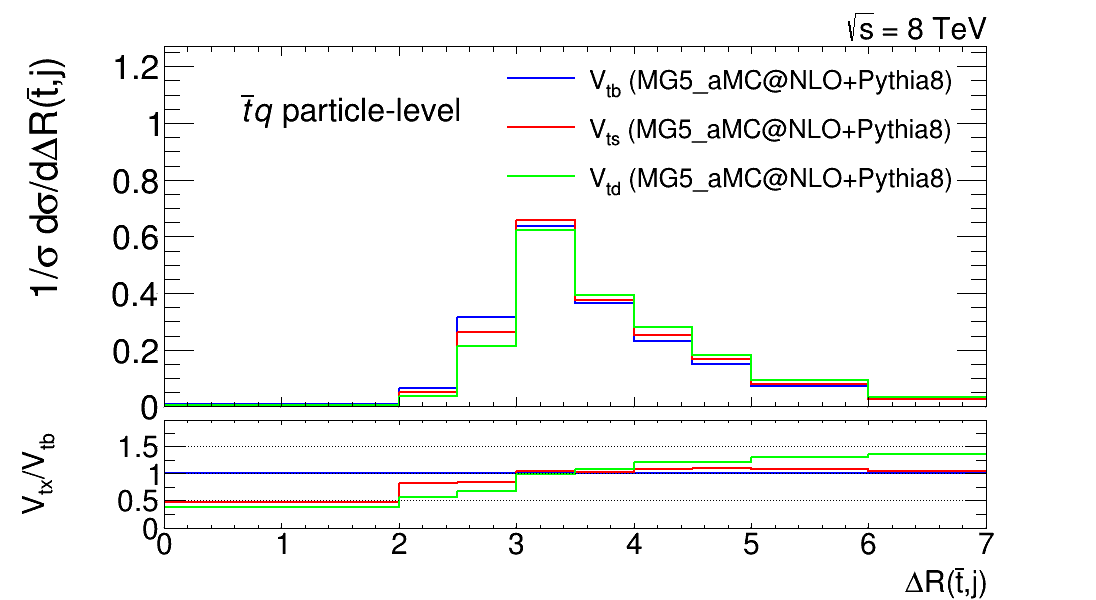}\\
    \end{tabular}
    \caption{Comparison of the normalized differential cross sections for single top (left) and anti-top (right) quark cross sections via Wtb (blue line), Wts (red line) and Wtd (green line) interactions at particle level as a function of the $b$-jet $|\text{y}|$ (first row) and  $\Delta$R (top, jet) (second row).
    \label{NEW-NormDiff}}
  \end{center}
\end{figure}

In addition to the aforementioned observables, it is found that the rapidity of the $b$-jet and the $\Delta$R = $\sqrt{\Delta\eta^2 + \Delta\phi^2}$ distance between the top quark and the forward jet are distinctive variables between Wtd and Wtb interactions. In Figure \ref{NEW-NormDiff}, normalized differential single top and anti-top quark production as a function of the $b$-jet rapidity and $\Delta$R(top, jet) are shown for Wtd, Wts and Wtb interactions.


\clearpage

\section{Global fit}
\label{fit}

The set of physical observables described in the  previous sections for probing the $\VTD$, $\VTS$ and $\VTB$  elements of the CKM matrix are measured by various experiments.
These elements of the CKM matrix can be most precisely determined, independently from the  unitarity of the CKM matrix, by a global fit to all available related measurements.

We construct a $\chi^2$ function in terms of the top CKM parameters according to the following definition:
\begin{equation}
\label{chi}
  \chi^2(\VTD,\VTS,\VTB) =  \sum_{\substack{i,j \text{ = bins or data points}} } \frac{(x_i^{th} - x_i^{exp})\rho_{ij}(x_j^{th} - x_j^{exp})}{\sigma_i \sigma_j}
\end{equation}
where $x_i^{th}$ is  the theoretical prediction, $x_i^{exp}$ is the experimental measurement and $\sigma_i$ is the corresponding uncertainty for the $i$-th observable; $\rho_{ij}$ is the correlation between
the  observables $i$ and $j$. We use the correlation matrices provided by the ATLAS experiment for the differential cross section measurements and set $\rho_{ij} = \delta_{ij}$ for other observables~\cite{Buckley:2015lku}.
Due to the fact that the single top inclusive cross sections that are used in this study are measured by different experiments and that the R parameter  is measured using top pair events, we expect that the effect of the   $\rho_{ij} = \delta_{ij}$ assumption would be negligible on the final results. On the other hand, the most effective observable in the fit is the single top $t$-channel differential cross section measurements in 69 bins in which bin-by-bin correlations are reported by the ATLAS  Collaboration and are fully considered on our results.
The uncertainties include experimental and theoretical uncertainties summed in quadrature, $\sigma_i=\sqrt{(\sigma_i^{th})^2+(\sigma_i^{exp})^2}$.

As experimental uncertainties we use $\pm$ one standard deviation as reported by the experiments. Asymmetric experimental errors are symmetrized, for simplicity, by conservatively treating the largest side uncertainty as a standard deviation. If experiments report  several sources of uncertainty the total error is obtained by adding them in quadrature.
Theoretical uncertainties originate from three sources, the choice of PDF, the choice of renormalization and factorization scales ($\mu_r$ and $\mu_f$) and the modeling of the parton shower and hadronization.
The  uncertainty  from  the  PDF  is  estimated  by reweighting the sample of
simulated single top $t$-channel events with respect to the  100  independent  replicas  of NNPDF31.
The  RMS  of  the  uncertainties  originating  from  the  variation  of all  replicas  is  taken  as  the  PDF uncertainty.
We estimate the uncertainties from variations in the renormalization and factorization scales  by reweighting the distributions of the single top production with different combinations of  $\mu_r$ and $\mu_f$ scales. The  central  values  of  the $\mu_r$ and $\mu_f$ scales correspond to the top quark mass and are varied independently over the range $m_t/2<\mu_{r,f}<2m_t$.
The uncertainty from parton shower is determined from independent simulated samples of the single top $t$-channel process in which the scales for initial and final state radiation is doubled and halved relative to their nominal values.
The statistical uncertainty due to the limited number of  simulated events is taken into account.

The best estimates of  the top related CKM elements are obtained by minimizing the total $\chi^2$ ($\chi^2_{\rm min}$) in a three dimensional phase space of $\VTD$, $\VTS$ and $\VTB$ .
The n-sigma uncertainty on the best estimates of the $\VTD$, $\VTS$ and $\VTB$ parameters is calculated by finding three dimensional contours which satisfy the following condition
\begin{equation}
  \chi^2(\VTD,\VTS,\VTB) =  \chi^2_{\rm min} + X
\end{equation}
where $X$ is  3.53, 7.82 and 11.3 for $1\sigma$, $2\sigma$ and $3\sigma$ contours, respectively.
We will report on the one dimensional confidence interval and on the best fit value, in two cases:
\begin{itemize}
\item Marginalized: the fit is performed with three free parameters ($\VTD$, $\VTS$, and $\VTB$) and then we integrate over two of the parameters;
\item Individual: the fit is performed with one free parameters ($\VTD$, $\VTS$, or $\VTB$) while the other two parameters are set to $\VTD$ = 0, $\VTS$ = 0, or $\VTB$ = 1.
\end{itemize}

\section{Results}
\label{result}

\begin{figure}[ht]
  \begin{center}
    \begin{tabular}{cc}
      \includegraphics[width=0.47\textwidth]{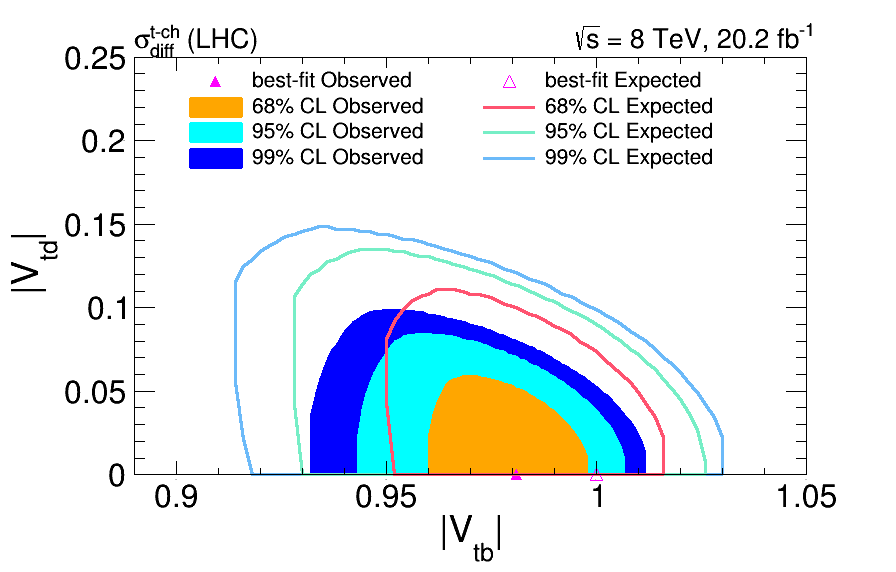}      &
      \includegraphics[width=0.47\textwidth]{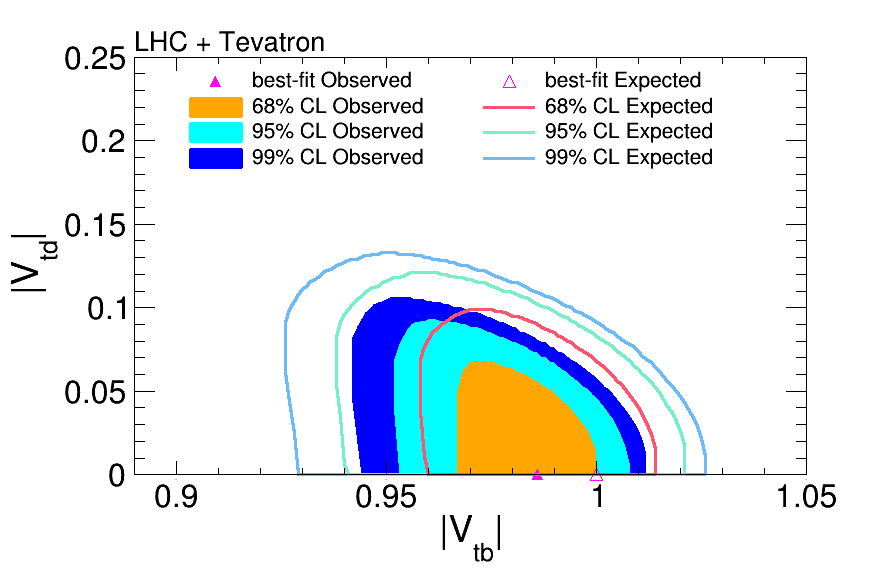} \\
      \includegraphics[width=0.47\textwidth]{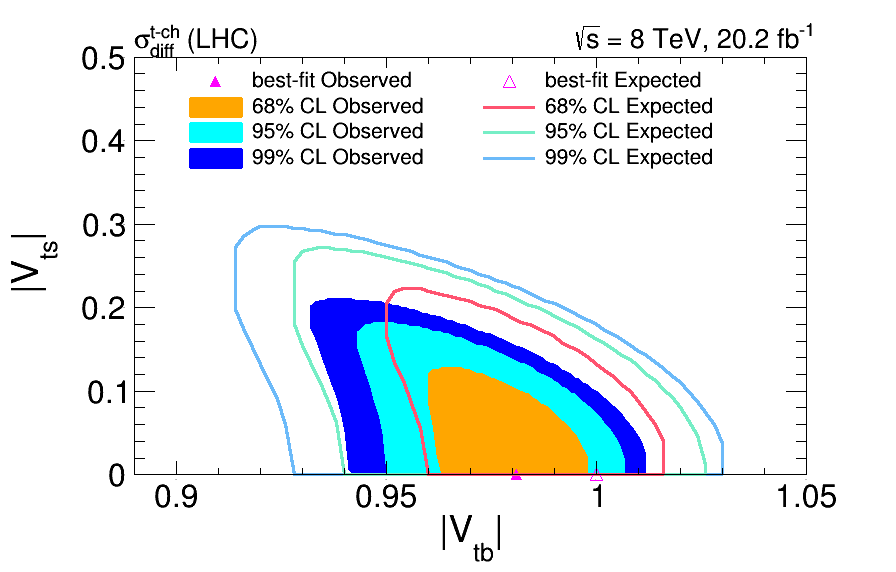}      &
      \includegraphics[width=0.47\textwidth]{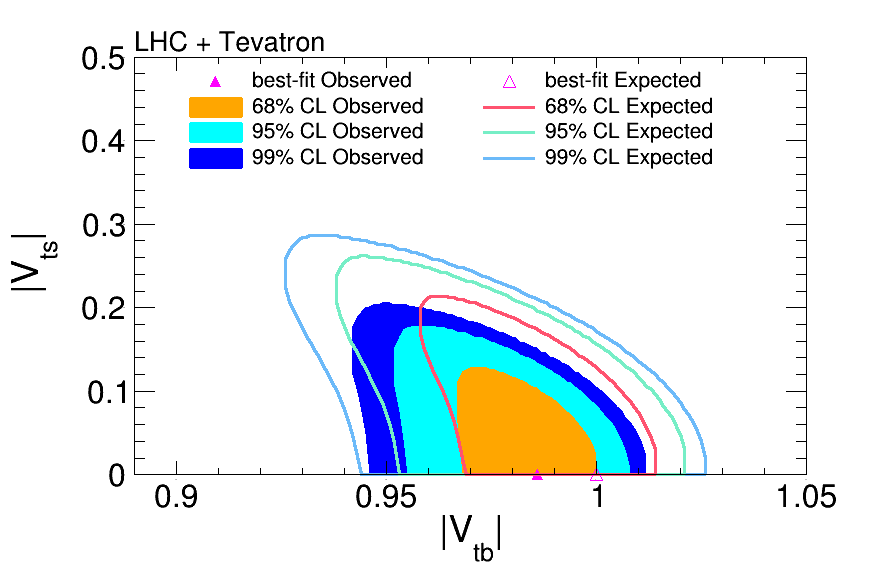} \\
      \includegraphics[width=0.47\textwidth]{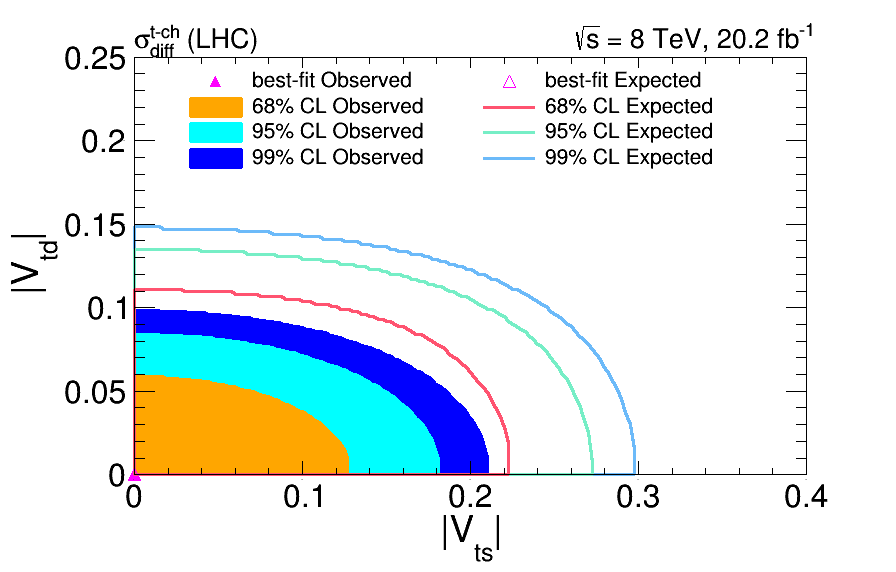} &
      \includegraphics[width=0.47\textwidth]{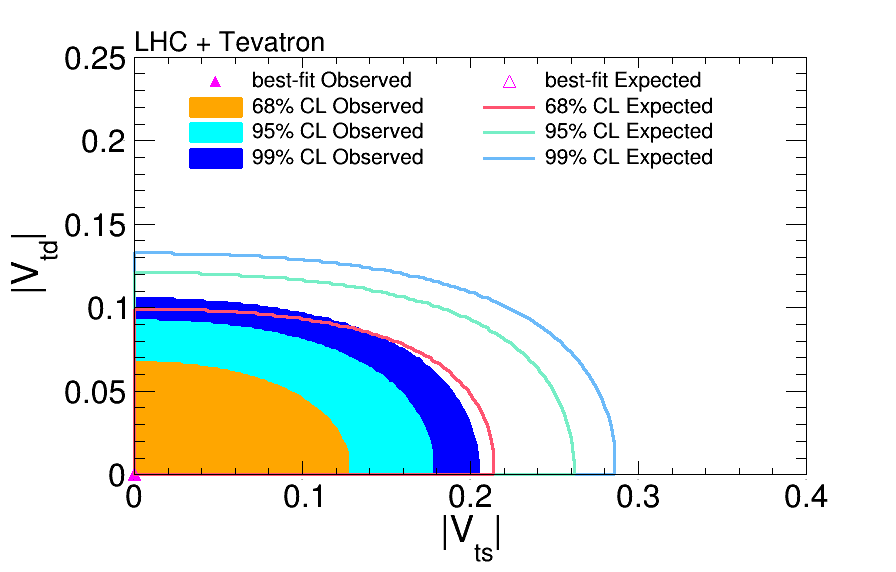} \\
      \end{tabular}
    \caption{The expected and observed best fit values together with the expected and observed 68\%, 95\% and 99\% confidence level regions in the ($\VTD$,$\VTS$), ($\VTB$,$\VTS$) and ($\VTD$,$\VTB$) planes from single top $t$-channel differential cross section measurements at 8 TeV by the ATLAS Collaboration \cite{Aaboud:2017pdi} in the left column  and from the combination of all LHC and Tevatron measurements (see Table \ref{exp-results}) in the right column.
    \label{compare2}}
  \end{center}
\end{figure}

In order to extract the best fit value and the uncertainty contours  on the $\VTD$, $\VTS$ and $\VTB$ parameters, we minimize the $\chi^2$ function defined in Equation \ref{chi}.
The simultaneous fit of the $\VTD$, $\VTS$ and $\VTB$ parameters is performed and the n-$\sigma$ allowed region is projected to the ($\VTD$,$\VTS$), ($\VTB$,$\VTS$) and ($\VTD$,$\VTB$) planes.
In this fit, the $\chi^2$  is calculated with respect to the $\VTD=0$, $\VTS=0$ and $\VTB=1$ point  of the phase space (expected) while the uncertainty is taken from the most precise available measurement for each observable.
The expected and observed best fit values together with 68\%, 95\% and 99\% confidence interval regions in the ($\VTD$,$\VTS$), ($\VTB$,$\VTS$) and ($\VTD$,$\VTB$) planes using single top $t$-channel differential cross section measurements at 8 TeV by the ATLAS Collaboration are shown in Figure~\ref{compare2} (left column). 
It is found that the constraints are improved by at least a factor of 2  when using the single top $t$-channel fiducial differential cross section measurements compared to the fiducial cross section measurements.
The fit is also performed with all the  LHC and Tevatron measurements noted in Table~\ref{exp-results} included, and results are shown in Figure \ref{compare2} (right column).

\begin{figure}[!ht]
  \begin{center}
        \includegraphics[width=0.75\textwidth]{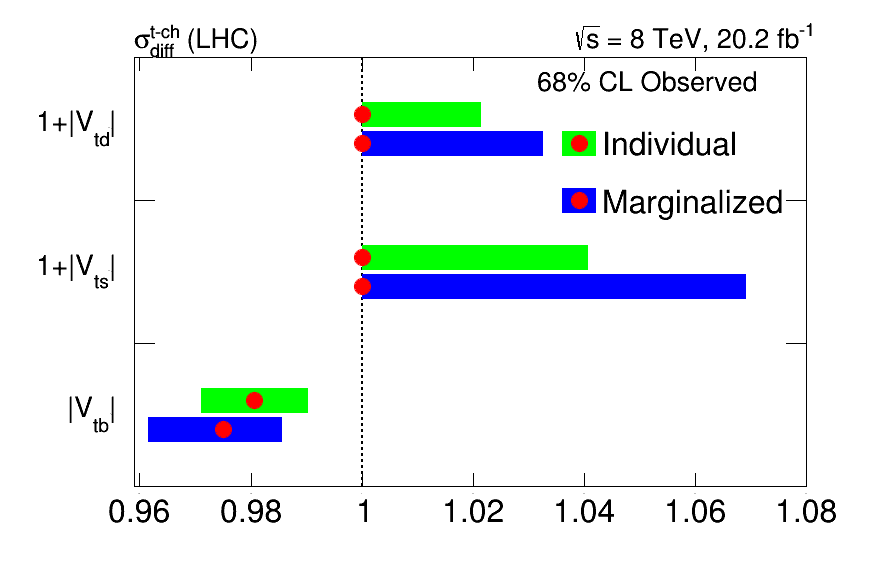}
      \includegraphics[width=0.75\textwidth]{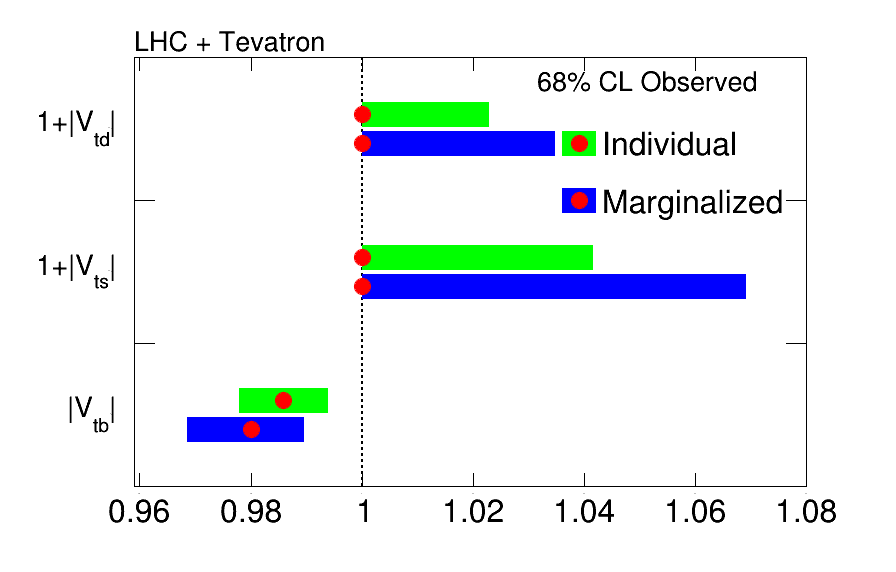}
    \caption{Observed individual and marginalized best fit values together with 68\% confidence interval for the $\VTD$, $\VTS$ and $\VTB$ CKM elements obtained from single top $t$-channel differential cross section measurements at 8 TeV (top) and from the combination of all LHC and Tevatron measurements (bottom).
    \label{fig:marg}}
  \end{center}
\end{figure}

The individual and marginalized best fit value and 68\% confidence interval for the $\VTD$, $\VTS$ and $\VTB$ parameters are summarized in Table \ref{FRVTX} and are shown in Figure \ref{fig:marg}.

The individual fit result on the $\VTB$ element can be interpreted as a single-parameter constraint on a generic VLQ model in which the $\VTD$, $\VTS$ and $\VTB$ CKM elements are scaled by a common multiplicative factor $k$, bound to $0 \leq k \leq 1$, leaving $R$ unchanged with respect to the SM expectation. In this scenario, $\VTD$ and $\VTS$ are practically unconstrained as the contribution of down- and strange-quark-initiated processes, already negligible in the SM, gets further reduced. Setting those contributions to zero for simplicity, the global fit result is $0.962 < k < 0.999$ at 95\% confidence level.

\begin{table}[!ht]
\centering
\caption{Observed individual and marginalized best fit values together with 68\% confidence interval for the $\VTD$, $\VTS$ and $\VTB$ CKM elements obtained from  single top $t$-channel differential cross section measurements at 8 TeV and from the combination of all LHC and Tevatron measurements (see Table \ref{exp-results}).}
\label{FRVTX}
\begin{tabular}{l|ll|ll}
\hline
\multirow{2}{*}{} & \multicolumn{2}{c|}{8 TeV differential}                       & \multicolumn{2}{c}{All combined}                             \\ \cline{2-5}
                  & \multicolumn{1}{l|}{marginalized} & individual                & \multicolumn{1}{l|}{marginalized} & individual                \\ \hline \hline
                  &                                   &                           &                                   &                           \\
$|V_{tb}|$        & $0.975^{+0.010}_{-0.014}$         & $0.981^{+0.010}_{-0.010}$ & $0.980^{+0.009}_{-0.012}$         & $0.986^{+0.008}_{-0.008}$ \\
                  &                                   &                           &                                   &                           \\
$|V_{ts}|$        & $0.000^{+0.069}_{0.000}$         & $0.000^{+0.041}_{0.000}$ & $0.000^{+0.069}_{0.000}$         & $0.000^{+0.041}_{0.000}$ \\
                  &                                   &                           &                                   &                           \\
$|V_{td}|$        & $0.000^{+0.032}_{0.000}$         & $0.000^{+0.021}_{0.000}$ & $0.000^{+0.038}_{0.000}$         & $0.000^{+0.023}_{0.000}$ \\
                  &                                   &                           &                                   &                           \\ \hline
\end{tabular}
\end{table}

\clearpage

\section{Prospects at the High-Luminosity LHC}
In this section, we present possible improvements in the direct $\VTQ$ measurement that  can  be achieved by the ATLAS or CMS experiments at the LHC for running conditions foreseen in the future.

It is planned that the LHC will deliver a total integrated luminosity of about 300 $\rm fb^{-1}$ and 3000~ $\rm fb^{-1}$ by the year 2023 and 2035, respectively~\cite{ATLAS:1502664,CMSCollaboration:2015zni}.
The instantaneous luminosity will increase substantially in the high luminosity LHC (HL-LHC), leading to more than 140 additional interactions per bunch crossing.   In addition, effect of radiation damage  could affect the sensitivity of the analyses. On the other hand,  improvements in the understanding of some of the systematic and theoretical uncertainty sources are expected thanks to the large data samples.

It was shown in Section \ref{result} that the most sensitive observables for constraining the $\VTQ$ elements are the differential cross section measurements of the single top $t$-channel process.
Thus, we only perform projections to higher integrated luminosities  using the differential observables  shown in Figure \ref{PL-Diff}. In order to find the theory prediction in each bin of the differential distributions, we use MC simulated samples for single top production at 13 TeV as  explained in Section \ref{fid}. The single top simulated samples for  $\VTD$, $\VTS$ and $\VTB$ are scaled to the cross section prediction at 13 TeV shown in Table \ref{xs}.
We use relative statistical and systematic uncertainties reported by the ATLAS Collaboration at 8 TeV \cite{Aaboud:2017pdi} to estimate uncertainties at 13 TeV in each bin of the differential distributions.
For the theoretical uncertainties, we use simulate samples at 13 TeV as  discussed in Section \ref{fit}.
We assume different scenarios for the systematic and theoretical uncertainties \cite{CMS-PAS-FTR-16-002,CMS-PAS-FTR-13-016}:
 \begin{itemize}
 \item 300 $\rm fb^{-1}$: All statistical and systematic uncertainties are taken from the ATLAS 8~TeV analysis \cite{Aaboud:2017pdi} and  scaled down by the square root of the integrated luminosity ratio ($\sqrt{20.2/300}$).  Theoretical uncertainties are found from simulated samples at 13 TeV.
 \item 3000 $\rm fb^{-1}$: All statistical and systematic uncertainties are taken from the ATLAS~8 TeV analysis \cite{Aaboud:2017pdi} and  scaled down by the square root of the integrated luminosity ratio ($\sqrt{20.2/3000}$).  Theoretical uncertainties are found from simulated samples at 13 TeV and scaled down by a factor 1/2.
 \end{itemize}

\begin{figure}[ht]
  \begin{center}
            \includegraphics[width=0.47\textwidth]{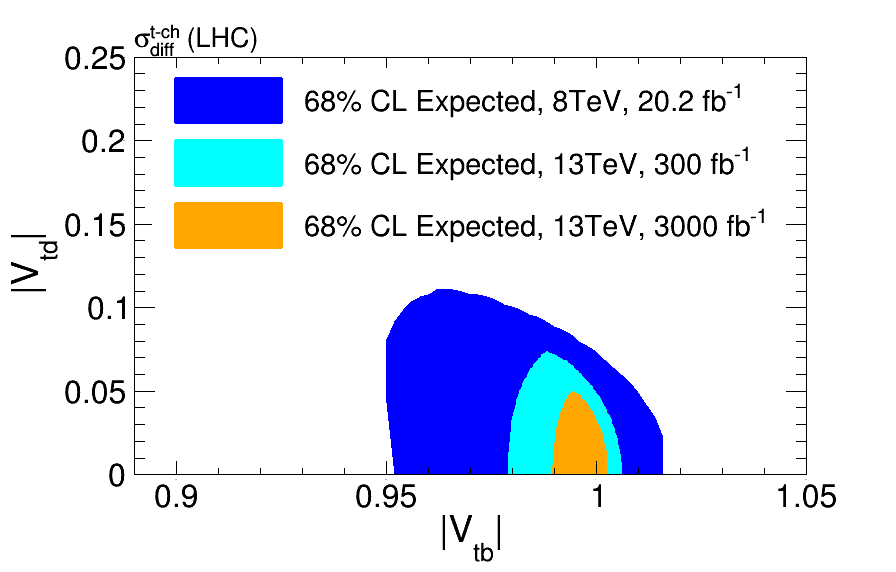}
            \includegraphics[width=0.47\textwidth]{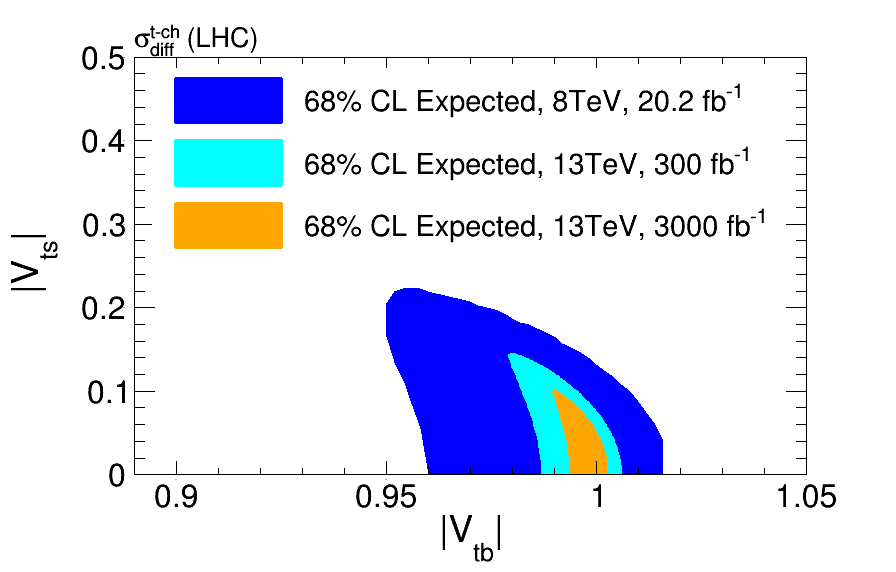}
            \includegraphics[width=0.47\textwidth]{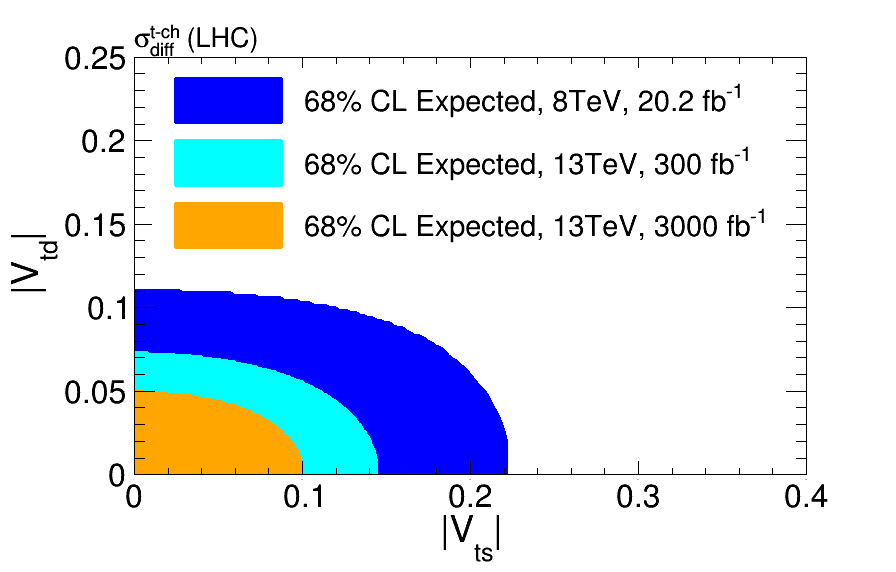}
    \caption{Comparison of the expected  68\% confidence level regions in the ($\VTD$,$\VTS$), ($\VTB$,$\VTS$) and ($\VTD$,$\VTB$) planes obtained using 8 TeV single top differential cross section measurements and  68\% confidence level regions expected for 300~\fb~ and 3000~\fb~ of data at 13 TeV.
    \label{proj2D}}
  \end{center}
\end{figure}

\begin{figure}[ht]
  \begin{center}
      \includegraphics[width=0.75\textwidth]{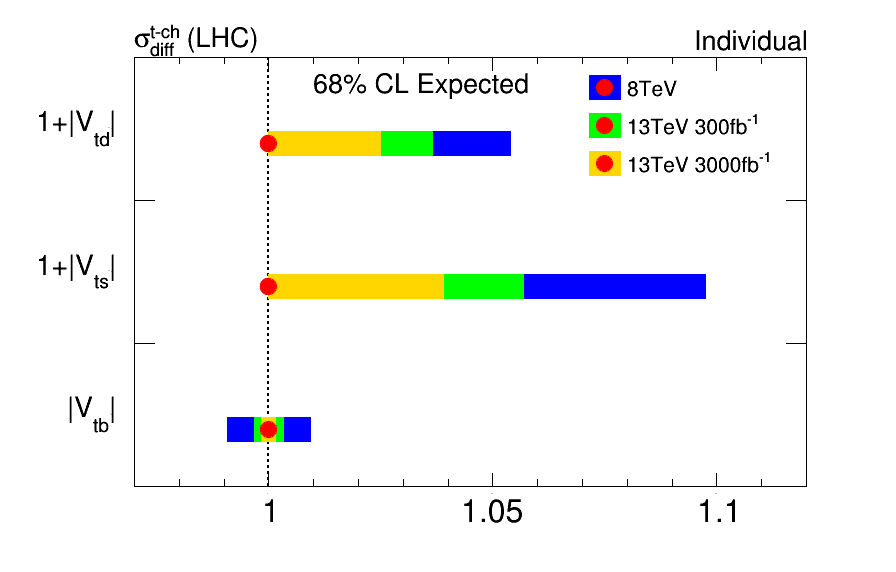}
      \includegraphics[width=0.75\textwidth]{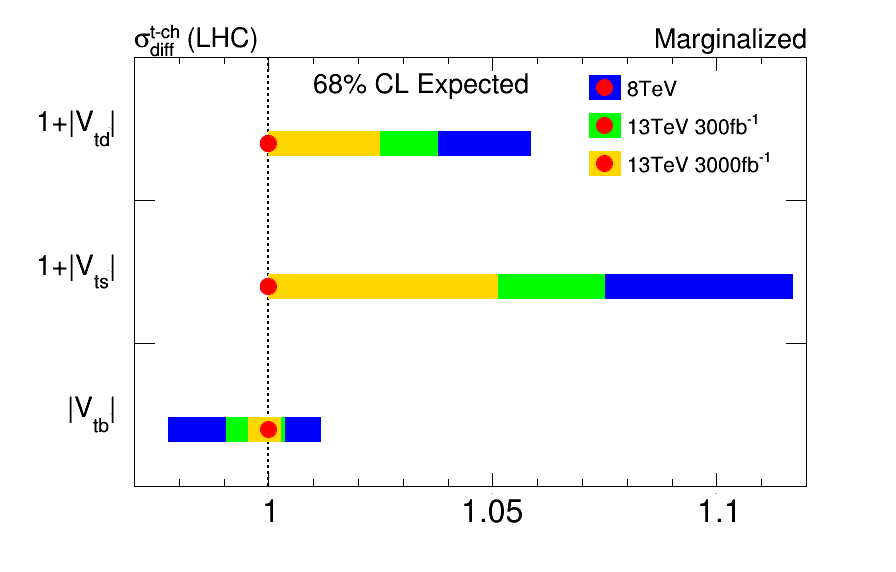}
    \caption{Projected 68\% confidence interval for individual (top) and marginalized (bottom) fits of the $\VTD$, $\VTS$ and $\VTB$ CKM elements from the differential cross section measurements of the single top production at 13 TeV LHC with 300 fb$^{-1}$ and 3000 fb$^{-1}$ of recorded data.  The  68\% confidence intervals  from the 8 TeV data are shown for comparison.
    \label{fig:prospects}}
  \end{center}
\end{figure}

In Figure \ref{proj2D}, the expected  68\% confidence level regions in the ($\VTD$,$\VTS$), ($\VTB$,$\VTS$) and ($\VTD$,$\VTB$) planes obtained using 8 TeV single top differential cross section measurements are compared to expected regions for  300~\fb and 3000~\fb of data at 13 TeV. Corresponding individual and marginalized   68\% confidence level regions for the $\VTD$, $\VTS$ and $\VTB$ CKM elements are shown in Table \ref{Projection} and Figure \ref{fig:prospects}.

\begin{table}[ht]
\centering
\caption{Expected 68\% confidence interval for individual and marginalized fits of the $\VTD$, $\VTS$ and $\VTB$ CKM elements from the differential cross section measurements of the single top production at 8 TeV LHC with 20.2 fb$^{-1}$ and 13 TeV LHC with 300 fb$^{-1}$ and 3000 fb$^{-1}$ of recorded data.}
\label{Projection}
\resizebox{\linewidth}{!}{%
\begin{tabular}{l|ll|ll|ll}
\hline
\multirow{2}{*}{} & \multicolumn{2}{c|}{8 TeV, 20.2 fb$^{-1}$}                       & \multicolumn{2}{c|}{13 TeV, 300 fb$^{-1}$} & \multicolumn{2}{c}{13 TeV, 3000 fb$^{-1}$}                                                         \\ \cline{2-7}
                  & \multicolumn{1}{l|}{marginalized} & individual                & \multicolumn{1}{l|}{marginalized} & individual  & \multicolumn{1}{l}{marginalized} & individual               \\ \hline \hline
                  &                                   &                           &                                   &                          & &\\
$|V_{tb}|$        & [0.9775,1.0115]         & [0.9907,1.0093]    & [0.9905,1.0035]  & [0.9968,1.0032]     & [0.9955,1.0025]& [0.9985,1.0015] \\
                  &                                   &                           &                                   &                          && \\
$|V_{ts}|$        & [0.0000,0.1171]         & [0.0000,0.0973]    & [0.0000,0.0750]  & [0.0000,0.0569]      & [0.0000,0.0510] & [0.0000,0.0390]  \\
                  &                                   &                           &                                   &                          && \\
$|V_{td}|$        & [0.0000,0.0585]        & [0.0000,0.0542]     & [0.0000,0.0377]  & [0.0000,0.0364]      & [0.0000,0.0247] & [0.0000,0.0249]\\
                  &                                   &                           &                                   &                          && \\ \hline
\end{tabular}}
\end{table}

Differential measurements for single top production in the $t$-channel are used in this projection study. Differential cross sections for single top production in the $tW$-channel are expected to be measured within good precision at HL-LHC \cite{Aaboud:2017qyi}.  This and other improvements on the R and top width measurements are expected to further improve the  precision attained in the future.
In addition to the higher precision in top quark measurements, all mentioned variables could be combined using a multi variate approach to maximize the sensitivity of the search.

\clearpage

\section{Summary}
Constraints on the CKM elements $\VTD$, $\VTS$ and $\VTB$ using processes involving top quarks without assuming unitarity of the CKM matrix have been set for the first time.
For this study we have combined the measurements of the  inclusive and differential cross section for single top quark production via $t$-channel and $tW$ production together with the measurements of  BR($t \rightarrow Wb$) and $\sigma_{\rm t-channel}^{\rm top} / \sigma_{\rm t-channel}^{\rm anti-top}$ from ATLAS, CMS, D0 and CDF experiments.
We have shown  the important role of the differential cross section measurements in the determination of the top related CKM elements. In addition to the top (anti-top) rapidity distributions, we found that the rapidity distribution of the forward associated jet in single top $t$-channel production is an important  observable for discriminating events from Wtd and Wts interactions with respect to the Wtb interaction.
We have also shown that the angular separation between the top quark and the forward jet, and the rapidity of $b$-jet in single top $t$-channel production are powerful variables  for constraining the $\VTD$, $\VTS$ and $\VTB$  CKM elements and we invite the LHC Collaborations to provide measurements of them.

The most important input for this study  is the differential measurements of $t$-channel single top-quark production using 20.2 fb$^{-1}$ of data collected by the ATLAS experiment in $pp$
collisions at 8 TeV at the LHC. The differential measurements for $tq$ and $\bar{t}q$ processes are performed in bins of the transverse momentum and the rapidity of the top quark and the forward jet, within the ATLAS fiducial cuts. We have used the ATLAS fiducial differential measurements to minimize model-dependent acceptance effects on the cross sections.

A global $\chi^2$  fit is performed on available experimental data to determine the $\VTD$, $\VTS$ and $\VTB$ simultaneously.  We have added  the modeling uncertainties on the theory to the reported experimental data. Correlation matrices between the experimental data points are included into the fit, when this information was provided by the experiment. After marginalizing on the other two top related CKM elements, we find   $|V_{tb}|$ = $0.980^{+0.009}_{-0.012}$, $|V_{ts}|$ = $0.000^{+0.069}_{0.000}$, $|V_{td}|$ = $0.000^{+0.038}_{0.000}$. We also performed a global $\chi^2$  fit on one of the CKM elements setting the other two elements to $\VTD$ = 0, $\VTS$ = 0 and $\VTB$ = 1, which gives $|V_{tb}|$ = $0.986^{+0.008}_{-0.008}$, $|V_{ts}|$ = $0.000^{+0.041}_{-0.000}$, $|V_{td}|$ = $0.000^{+0.023}_{0.000}$.

In the near future, LHC is expected to collect large datasets at $\sqrt{s}$ = 13 TeV. In addition to the reduction of the statistical uncertainties, possible improvements in the understanding of systematic uncertainties are expected.  We have projected the results based on the ATLAS differential cross section measurements at $\sqrt{s}$ = 8 TeV to larger datasets of 300 and 3000 fb$^{-1}$ at $\sqrt{s}$ = 13 TeV. Compared to the 8 TeV results, the precision on the $\VTD$, $\VTS$ and $\VTB$ measurements could be improved  by a factor of $\sim$~2.5 ($\VTB$), ~1.5 ($\VTS$, $\VTD$) for 300 fb$^{-1}$ data and $\sim$~5($\VTB$), ~2($\VTS$, $\VTD$) for 3000 fb$^{-1}$ data, respectively.
\section*{Acknowledgments}
The authors wish to thank Wolfgang Wagner for clarifying several details of the ATLAS differential cross section analysis~\cite{Aaboud:2017pdi}, including the need to correct for the $W$-boson branching ratio into leptons. We also thank Pascal Vanlaer and Fabio Maltoni for their useful comments on this study. This work was partly supported by F.R.S.--FNRS under the ``Excellence of Science -- EOS'' --be.h project n. 30820817.

\bibliography{vtx}{}
\bibliographystyle{JHEP}
\end{document}